\documentstyle[12pt,epsf]{article}
\def\xd{x_D}
\def\yd{y_D}
\def\zd{z_D}

\def\xba{\overline}

\newcommand{\bfc}{{\cal M}}
\newcommand{\bfd}{{\bf d}}
\newcommand{\bfdd}{{\bf D}}
\newcommand{\nup}{{\tilde N^{3\sigma}_{B}}}

\newcommand{\lsim}{\mathrel{\lower4pt\hbox{$\sim$}}
\hskip-12.5pt\raise1.6pt\hbox{$<$}\;}

\newcommand{\gsim}{\mathrel{\lower4pt\hbox{$\sim$}}
\hskip-12.5pt\raise1.6pt\hbox{$>$}\;}

\def\beq{\begin{equation}}
\def\eeq{\end{equation}}
\def\beqa{\begin{eqnarray}}
\def\eeqa{\end{eqnarray}}
\def\shat{\hat s}

\begin{document}
\begin{flushright}
AMES-HET-00-09 \\
BNL-HET-00/31\\
\end{flushright}
\medskip

\begin{center}
{\bf 
Improved Methods for Observing CP Violation in $B^\pm\to KD$
and Measuring the CKM Phase $\gamma$\\

}

\medskip

David Atwood$^a$, 
Isard Dunietz$^b$ 
and
Amarjit Soni$^c$
\end{center}
\medskip

\begin{flushleft}
$a$) Dept. of Physics, Iowa State University, Ames, IA 50011 \\
$b$) Fermilab, Batavia IL\ \  60500\\
$c$) Theory Group, Brookhaven National Laboratory, Upton, NY\ \ 11973
\end{flushleft}
\medskip

\begin{quote}
{\bf Abstract:} 

Various methods are discussed for obtaining the CKM angle $\gamma$
through the interference of the charged $B$-meson decay channels $B^-\to
K^- D^0$ and $B^-\to K^- \xba D^0$ where the $D^0$ and $\xba D^0$ decay to
common final states.  It is found that choosing final states which are not
CP eigenstates can lead to large direct CP violation which can give
significant bounds on $\gamma$ without any theoretical assumptions.  If
two or more modes are studied, $\gamma$ may be extracted with a precision
on the order of $\pm 15^\circ$ given $\sim 10^8$ $B$-mesons.  We also
discuss the case of three body decays of the $D$ where additional
information may be obtained from the distribution of the $D$ decay
products and consider the impact of $D\xba D$ oscillations. 

\end{quote}
\newpage

\section{Introduction}\label{section_1}

The only manifestations of CP violation observed to the present time are
those in the neutral kaon system. The advent of several machines capable
of producing large number of $B$ mesons makes it likely that at least some
examples of CP violation in $B$ physics will soon be seen. 

$B$ factories at SLAC and KEK have been specifically constructed to
observe CP violation in the time dependent oscillation of $B^0\xba B^0$
when $B^0$ decays to a CP eigenstate such as $\psi K_s$ or $2\pi$. In such
experiments one has the advantage that the CP violating phase may be in a
few cases cleanly extracted. In contrast direct CP violation which can
occur in $B^\pm$ decay always appears in combination with a strong phase
which cannot be easily determined since it has its origin in strong
interaction physics. In order to put the extraction of the CP odd phase on
the same footing as the clean oscillation experiments it is thus important
that some way of eliminating the uncertainties due to the strong phase be
found.

Direct CP violation in $B^\pm$ decay may prove to be an important
component of the future $B$ physics program because it offers signals
sensitive to the angle $\gamma$ of the unitarity triangle~\cite{utriang}. 
Here we focus on effects which originate from the interference of $B^-\to
K^- D^0$ with $B^-\to K^- \xba D^0$. In contrast, experiments involving
$B^0$ may only extract $\alpha$ (from $B^0\to 2\pi$ for instance) and
$\beta$ (from $B^0\to \psi K_S$).  The unitarity triangle which follows
from the unitarity of the CKM matrix is a key prediction of the Standard
model. Independent measurements to over determine the triangle by
measuring each of its sides and angles therefore provide a non-trivial
test of the standard model.  Furthermore, the Standard Model predicts that
a number of different measurements of CP violation will depend on the same
phase $\gamma$ (e.g. direct CP violation in $B\to K\pi$ and oscillations
in $B_s$).  Comparing the results of these experiments will thus be a
sensitive test for new physics.

Since the modes we consider are direct and not time dependent, they may be
observed in any experimental setting where large numbers of $B$ mesons are
produced. Aside from the SLAC and KEK asymmetric $B$-factories, these
include CLEO and hadronic $B$ experiments such as BTeV, CDF, D0 and LHCB
or a high luminosity $Z$ factory.

A method for putting the determination of $\gamma$ on the same footing as
the oscillation experiments by using direct CP violation was first
suggested in \cite{gronauwyler} where the decay $B^-\to K^-D^0$ is
considered followed by the decay of $D^0$ to a CP eigenstate (see also
\cite{twoB}).  In this case the fact that the final state is a CP
eigenstate means that it will interfere with the channel $B^-\to K^-\xba
D^0$.  In Section~2, for self containment, we will review the important
features of this method and discuss some problems which arise in its
implementation.

These problems can be resolved through a more general method using non-CP
eigenstates suggested in~\cite{twoC,threeA} and refined in~\cite{ads}
which not only enable a clean extraction of $\gamma$ but in addition have
the attractive feature that they can give rise to large CP asymmetries.
This is outlined and expanded upon in Section~3 and 4. In Section~5 we
estimate some relevant branching ratios and obtain a rough estimate of the
attainable accuracy in extracting $\gamma$.

In Section~6 we discuss the somewhat more general case where the
distributions of three body decays of $D^0$ are considered. Here we
consider two different methods for obtaining $\gamma$. First, one can fit
the distributions to a resonant channel model where each of the channels
can be considered a quasi two body mode. The phases between the channels
will thus give additional constraints on the fit. Second, if each point on
the Dalitz plot is considered a separate mode, in some cases the
accumulated inequality bounds on $\sin^2\gamma$ can provide a
determination of $\sin^2\gamma$.  In section~7 we give our summary and
conclusions. 

Throughout this paper we will assume that the effects of $D\xba D$ mixing
are negligible. In an appendix, we discuss the impact of $D\xba D$ mixing
on this method for determining $\gamma$. Specifically, if one uses the
time interval between the parent $B^-$ decay and the subsequent $D$
decay, one can eliminate the effects of mixing. One can also obtain
information about $\gamma$ in time independent studies.

\section{Using CP Eigenstates Decays of $D^0$}\label{section_2}

Gronau London and Wyler's
strategy (GLW) to extract $\gamma$ from CP violation in the decay 
$B^-\to K^-
D^0$ followed by $D^0\to$~CP~eigenstate is to separately determine
the 
branching ratios~\cite{gronauwyler}:

\begin{itemize}
\item[(a)]
$Br(B^-\to K^- D^0)$
\item[(b)]
$Br(B^-\to K^- \xba D^0)$
\item[(c)]
$Br(B^-\to K^-  D^0_{+})$
or
$Br(B^-\to K^-  D^0_{-})$
\end{itemize}

\noindent together with their conjugates where $D^0_\pm$ denote the CP
eigenstate $D^0_\pm=(D_0\pm \xba D_0)/\sqrt 2$.

Once $a$, $b$, $c$ and the corresponding quantities for the conjugate
modes are known then one can separate out the interference effects and
thus determine $ \cos(\zeta_K+\gamma)$ and $\cos(\zeta_K-\gamma)$
simultaneously where $\zeta_K$ is the CP conserving strong phase
difference between $B^-\to K^- D^0$ and $B^-\to K^- \xba D^0$ while
$\gamma$ is the CP violating weak phase difference.  From these two
cosines, the values of the actual phase angles $\zeta_K$ and $\gamma$ can
clearly be determined, up to a eight-fold discrete ambiguity as will be
discussed in more detail in section~\ref{section_4}.

The rate for (c) is experimentally observed through the decay to a CP
eigenstate such as $K^+K^-$, $K_s\phi$ and $\pi^+\pi^-$ and presents no
problem in principle.  Likewise the decay (a) should be readily measurable
through either leptonic or hadronic modes of the $D^0$. As it stands,
however, this method has a very serious problem in measuring the branching
ratio of (b), $Br(B^- \to K^-\xba D^0)$.  The detector must distinguish
$\xba D^0$ from $D^0$ to determine the decay rate to this rare mode. The
background decay $Br(B^- \to K^- D^0)$ (i.e. (a)) is 
expected to be larger by about two orders of magnitude.

There are only two ways that could possibly be used to 
tag the flavor of the $\xba D^0$:
\begin{itemize}
\item[(a)] through semi-leptonic decays
\item[(b)] through hadronic decays
\end{itemize}
We will consider each of these individually and show that neither is 
fully satisfactory.

The semi-leptonic tag, i.e. the quark level process $\xba c \to \ell^-
\xba s \xba \nu_\ell$, has the problem that there is an overwhelming
background from the direct semileptonic decay of the $B$ meson. The signal
here is the sequential decay: 

\beq
B^- \to K^-\xba D^0 \quad ; \quad 
\xba D^0 \to \ell^- \xba \nu_\ell X_{\xba s} 
\label{btokd}
\eeq

\noindent while the background is from the direct semi-leptonic decay of the
$b$ quark followed by hadronic decays of the charm quark:

\beq 
B^- \to \ell^-\xba \nu_\ell X_c \quad ; \quad c\to s u\xba d
\label{btoell} 
\eeq

\noindent Both give rise to the same sign lepton and while there are
several features distinguishing the signal from the background it
represents a serious problem as the branching ratio for the signal is
expected to be ${\cal O}(10^{-6})$ whereas that for the background is
expected to be ${\cal O}(10^{-1})$. The signal/background ratio is thus
dauntingly small.

Let us note some of the characteristics of the signal which distinguish it
from the background. First of all, in the $B^-$ rest frame the $K^-$ tends
to be monochromatic.  Also the signal tends to yield a pair of kaons
$K^-K^+$ (or $K^-K^0$) whereas the background dominantly has one kaon,
$K^-$ or $\xba K^0$. In the signal the semi-leptonic decays of the $\xba
D^0$ originate from a tertiary vertex in sharp contrast to the case of the
background. It is difficult to say whether these distinctions are enough
to separate the signal.  Detector specific studies are required to give
reliable answers but the size of the signal/background ratio does not give
us much ground for optimism.

While the use of such a leptonic tag is likely to be impractical, the
hadronic decay has a more fundamental problem. In this case one would
detect the decay $B^-\to K^-\xba D^0$ by observing a final state where the
$\xba D^0$ decays through a Cabibbo-allowed (CBA) process, for instance
$\xba D^0 \to K^+\pi^-$.  Unfortunately, the doubly-Cabibbo suppressed
(DCS) decay of $D^0$ (e.g. $D^0\to K^+\pi^-$) will also lead to the same
final state at a branching ratio two orders of magnitude smaller.

On the other hand, as was pointed out in~\cite{ads}, the initial
decay $B^-\to K^-\xba D^0$ is color-suppressed (CLS) while the decay
$B^-\to K^-D^0$ is color-allowed (CLA). Thus difference in the production
rate tends to offset the difference in the decay rates for the two
processes and since the final states are really indistinguishable, they
will interfere quantum mechanically.

As a specific example consider the possible tag $\xba D^0\to K^+\pi^-$. 
The signal from $B^-\to \xba D^0 K^-$ and the background from $B^-\to D^0
K^-$ will be given by the sequences:

\beqa
A)~ B^-\to K^- \xba D^0   & ; & \quad \xba D^0 \to K^+\pi^-
\label{eqasix} \\
B)~ B^-\to K^- D^0        & ; & \quad D^0 \to K^+ \pi^- 
\label{eqaseven}
\eeqa

These two decay chains have the qualitative form.

\beqa
A: & & {CLS} \otimes {CBA} \label{eqaeight} \\
B: & & {CLA} \otimes {DCS} \label{eqanine} 
\eeqa

\noindent Numerically the ratio of these two amplitudes is appreciably
close to 1 since it is expected to be given by:

\begin{eqnarray} 
\left| \frac{M(B^-\to K^- D^0 [\to K^+\pi^-])}{M(B^-\to K^-\xba D^0
[\to K^+\pi^-])} \right|^2 
\approx \left| \frac{V_{cb}V^\ast_{us}}{V_{ub}
V^\ast_{cs}} \right|^2 \left| \frac{a_1}{a_2} \right|^2 \frac{B(D^0\to
K^+\pi^-)}{B(\xba D^0\to K^+\pi^-)} \label{eqaten}
\end{eqnarray}

\noindent Here $a_1$ and $a_2$ control the relative sizes of the CLA
and the CLS amplitudes. 
Experimentally~\cite{aratio} the indications are that

\beq
\left| \frac{a_2}{a_1} \right | \approx  0.26 \pm 0.07 \pm 0.05
\label{eqaeleven}
\eeq

\noindent which roughly agrees with the simple color counting value
of $1/3$.

Let us consider $B(D^0\to K^+\pi^-)/B(\xba D^0 \to K^+\pi^-)$ which is of
order $\lambda^4$.  We can formulate an estimate taking into account SU(3)
breaking effects in the form factors and the decay constants. For the
present purpose we use the experimental result~\cite{pdb}:

\beq
\frac{Br(D^0 \to K^+\pi^-)}{Br(\xba D^0\to K^+\pi^-)} = .0072\pm .0025 
\label{eqatwelve}
\eeq

\noindent as well as the ratio of CKM elements~\cite{pdb}:

\beq
\left| \frac{V_{ub}}{V_{cb}} \right| = .08\pm.02.   
\label{eqathirteen} 
\eeq

\noindent 
Then with
$\lambda=0.23$ and the central values from
(Eqns.~\ref{eqaeleven}--\ref{eqathirteen}) we therefore get:

\beq
\left |
\frac{M(B^-\to K^- D^0 [ \to K^+\pi^-])}{M(B^- \to K^- \xba D^0 [\to
K^+\pi^-] )}
\right |^2
\approx 1 \label{eqafourteen}
\eeq

\noindent The two amplitudes are roughly comparable and we cannot tell
whether the charmed particle was a $D^0$ or a $\xba D^0$. $B^-\to K^-\xba
D^0$ with a $\xba D^0$ decaying hadronically will give rise to a final
state which is indistinguishable from the corresponding decay of the $D^0$
in $B^-\to K^- D^0$. The two amplitudes will thus be subject to large
quantum mechanical interference effects.  The use of a hadronic tag for
determining $B(B^-\to K^-\xba D^0)$ for the GLW method appears therefore
to be ruled out.

Despite
these difficulties, one need not discard the GLW
approach.  The only input which is lacking is the branching ratio
$B^-\to K^- \xba D^0$.  It may be possible to theoretically estimate this
quantity which will allow the GLW program to go forward.

Here we consider
an alternative approach where we take advantage of these interference
effects to enhance CP violation and ultimately provide another way for a
clean (i.e. no penguin pollution) way of extracting $\gamma$. 
As discussed in~\cite{ads} the fact that these two amplitudes have large
interference effects implies that there will be large CP violating
asymmetries in such combined decay channels which in turn gives us a
handle on measuring $\gamma$. Thus, it is instructive to consider why CP
violation will be enhanced in this case as compared to GLW.

In the eigenstate case, the size of the expected CP asymmetry is
controlled only by the ratio of the amplitudes for $B^-\to K^- D^0$ versus
$B^-\to K^-\xba D^0$.  Following the above estimate and taking into
account the appropriate CKM factors:

\beq
\left| \frac{M(B^-\to K^-\xba D^0)}{M(B^-\to K^-D^0)} \right| \sim
\left| \frac{a_2}{a_1}\right| \cdot
\left|\frac{V_{ub}}{V_{cb}V^\ast_{us}} \right| \simeq 0.1
\label{eqafifteen}
\eeq

\noindent This means that the maximum possible size of the CP asymmetry is
expected to be 0(10\%).  In contrast eq.~(\ref{eqafourteen}) implies that
the two interfering amplitudes have roughly the same magnitude and so the
interference effects, and in particular CP violation, will be near
maximal.

Of course the search for large direct CP violating signals is interesting
in its own right but the real goal is to extract the angle $\gamma$ from
experimental results cleanly, that is without any reference to a model for
hadronization. While the original idea of~\cite{gronauwyler}, though sound
in principle is probably not practical, the basic concept may be used if
one observes two or more distinct hadronic states that are common decay
products of $D^0$ and $\xba D^0$ (as all hadronic final states of $D^0$
are).  With this information, one can reconstruct $\gamma$ cleanly.  We
now consider a number of methods based on this idea.

\section{Non-CP Eigenstates Decays of $D^0$}\label{section_3}

Using this idea, let us now consider the case where the two channels

\begin{eqnarray}
B^-\to K^- D^0 ~~~~
B^-\to K^- \xba D^0
\label{eqsixteen}
\end{eqnarray}

\noindent interfere because both $D^0$ and $\xba D^0$ decay to some common
final state $X$.  In the GLW method the specific case where $X$ is a CP
eigenstate such as $K_s\pi^0$ was chosen while we will focus on the
instance, considered in~\cite{ads}, where $X$ is not a CP eigenstate.  In
particular, following the logic of the previous section, the case where
$D^0\to X$ is a DCS decay and $\xba D^0\to X$ is a CBA decay is of
particular interest, for instance $X=K^+\pi^-$.

In order to formulate an expression for the rates, let us define the
following branching ratios:

\begin{eqnarray}
a(k)=Br(B^-\to k^- D^0) &~& \xba a(k)=Br(B^+\to k^+ \xba D^0)\nonumber\\
b(k)=Br(B^-\to k^- \xba D^0) &~& \xba b(k)=Br(B^+\to k^+ D^0)\nonumber\\
c(X)=Br(D^0\to X) &~&  \xba c( X)=Br( \xba D^0\to X) \nonumber\\
c(\xba X)=Br(D^0\to \xba X) &~&  \xba c( \xba X)=Br( \xba D^0\to \xba X) 
\nonumber\\ 
d(k,X)=Br(B^-\to k^- [X]) &~&
\xba d(k,\xba X)=Br(B^+\to k^+ [\xba X]) 
\nonumber\\
&\ &
\end{eqnarray}

Here $k^\pm$ represents either $K^\pm$ or $K^{*\pm}$ (or indeed one may
consider any other kaonic resonance or system of strangeness=$-1$ and well
defined CP) and [$X$] is the common $D^0$ and $\xba D^0$ decay channel
observed. Thus the combined rates $d(k,X)$ and $\xba d(k,\xba X)$ include
the effects of the interference of the two channels.

In the standard model, it is expected that $a(k)=\xba a(k)$, $b(k)=\xba
b(k)$ and $\xba c(X)= c(\xba X)$ all of which we will assume from here on. 
In general, however if $\gamma\neq 0$ one expects $d(k,X)\neq\xba d(k,\xba
X)$ (CP violation) and indeed the value of the quantities $d$, $\xba d$
may be expressed in terms of $a$, $b$ and $c$ as:

\begin{eqnarray}
d(k,X)&=& a(k)c(X)+b(k)c(\xba X)
\nonumber\\
&&+2\sqrt{a(k)b(k)c(X)c(\xba X)}\cos(\zeta_k+\delta_X+\gamma)
\nonumber\\
\xba d(k,\xba X)&=& a(k)c(X)+b(k)c(\xba X)
\nonumber\\
&&+2\sqrt{a(k)b(k)c(X)c(\xba X)}\cos(\zeta_k+\delta_X-\gamma)
\label{eqnd}
\end{eqnarray}

\noindent
where $\zeta_k$ is the strong phase difference between $B^-\to k^- D^0$ and
$B^-\to k^- \xba D^0$; $\delta_X$ is the strong phase difference between $D\to
X$ and $D\to \xba X$ and $\gamma$ is the CP violating weak phase difference
between $B^-\to k^- D^0$ and $B^-\to k^- \xba D^0$.

In the standard model $\gamma$ is given directly from the CKM elements 

\begin{eqnarray}
\gamma=arg(-V_{ud}V_{ub}^*V_{cb}V_{cd}^*).
\end{eqnarray}

\noindent Existing data does not constrain $\gamma$ very much giving an
allowed range in the SM at 95\% c.l. \cite{gammaref} of $36^\circ \leq
\gamma \leq 97^\circ$ corresponding to $0.35\leq \sin^2\gamma$.

The strong phases $\zeta_k$ and $\delta_X$ that result from hadronic final
state interactions cannot be reliably calculated with any known method and
must be determined experimentally.  Here we will take the approach that
information about $\delta_X$ and $\zeta_k$ are extracted from the data
along with $b(k)$ and $\gamma$.

The above may be made somewhat more general by considering the class of
modes $B^-\to k^- \bfdd$ where $\bfdd$ is an excited $D$ meson.  Let us
suppose that $\bfdd$ subsequently decays $\bfdd\to D+N^0$ where
$N^0$ is a single particle and the $D$ then decays into a CP
non-eigenstate mode of the type we have considered above. For instance 
$\bfdd$ may be a $D^{*0}$ with $k^-=K^-$ and we may use either of the 
following decay chains: 

\begin{eqnarray}
B^-&\to& K^-(D^*\to\pi^0[D\to X])\nonumber\\ 
B^-&\to& K^-(D^*\to\gamma[D\to X])
\end{eqnarray}

\noindent Here $D^0\to X$ is a DCS mode. In these sort of examples the
analysis would be essentially the same as considered for the ground state
of the $D$. The only constraint on this generalization is that one of $\{
k,\bfdd\}$ should have spin 0 so that there is only one partial wave
otherwise multiple partial waves would have to be separated and the
analysis for the extraction of $\gamma$ would be more complicated and
has been considered in~\cite{sings}.

\section{Methods for Extraction of $\gamma$}\label{section_4}

Let us now turn our attention to the extraction of the weak phase $\gamma$
and incidentally also the total strong phase $\xi(k,X)=\zeta_k+\delta_X$. 
We will consider two scenarios under which such a reconstruction is
possible assuming that the values of $c(X)$ and $c(\xba X)$ have been
determined in advance and that the rate of $D^0 \xba D^0$ mixing is
negligible.  In the appendix we show how the possible effects of mixing
may be removed from the data.

\bigskip

{\it {\bf\it Case 1:} 
For one particular mode, $(k,X)$, $d(k,X)$ and $\xba d(k,\xba X)$ are known
and in addition $a(k)$ and $b(k)$ are known. 
}

\bigskip

This is the simple generalization of the GLW method to the case where $X$
is not a CP eigenstate. We thus assume that the branching ratio $b(k)$ may
be obtained from tagging the $\xba D$ with the semileptonic decays as
discussed in section~\ref{section_2} or as the result of some theoretical
estimate~\cite{yamamoto}.  If $b(k)$ can be obtained in some way,
this method is less involved than the other method we will discuss below.

As in the GLW method we are required to extract the interference terms in
$d$ and $\xba d$ by solving the two equations (\ref{eqnd}) which we can 
rewrite: 

\begin{eqnarray}
d(k,X)&=&
\left[a(k)c(X)+b(k)c(\xba 
X)\right]\left[1+R(k,X)\cos(\zeta_k+\delta_X+\gamma)\right] \nonumber\\
\xba
d(k,\xba X)&=&\left [ a(k)c(X)+b(k)c(\xba 
X)\right ]
\left[ 1+R(k,X)\cos(\zeta_k+\delta_X-\gamma)\right]. 
\nonumber\\
\label{onemode}
\end{eqnarray}

\noindent
where $R(k,X)$ is 

\begin{eqnarray}
R(k,X)= { 2\sqrt{a(k)b(k)c(X)c(\xba X)} \over  a(k)c(X)+b(k)c(\xba X) }
\label{rdef}
\end{eqnarray}

Defining 

\begin{eqnarray}
\lambda_1 &=& {1\over R(k,X)} \left[ {{d(k,X)\over a(k)c(X)+b(k)c(\xba
X)}-1}\right] \nonumber\\
\lambda_2 &=& {1\over R(k,X)} \left[ {{\xba d(k,X)\over a(k)c(X)+b(k)c(\xba
X)}-1}\right], \label{lambdadef}
\end{eqnarray}

\noindent the solution for $\gamma$ and $\xi(k,X)=\zeta_k+\delta_X$ is:

\begin{eqnarray}
\gamma  &=& {1\over 2}(\sigma\cos^{-1} \lambda_1 -\tau\cos^{-1}\lambda_2)
+n\pi; 
\nonumber\\
\xi(k,X) &=& 
{1\over 2}(\sigma\cos^{-1}\lambda_1 + \tau\cos^{-1} \lambda_2)
+n\pi.
\label{solution_a}
\end{eqnarray}

In the above $\sigma$, $\tau\in\{\pm 1\}$ and $n\in\{0,1\}$ expressing the
fact that there is a eight fold ambiguity~\cite{ads_cross_glw} since the
sign in front of each of the $\cos^{-1}$ functions is undetermined and we
can add $\pi$ simultaneously to $\xi$ and $\gamma$ without changing the
results.  Specifically, the eight solutions giving results identical to a
given $(\xi,\gamma)$ are $\{ (\xi,\gamma)$,
$(-\xi,-\gamma)$,$(\gamma,\xi)$, $(-\gamma,-\xi), (\pi+\xi,\pi+\gamma)$,
$(\pi-\xi,\pi-\gamma)$, $(\pi+\gamma,\pi+\xi)$, $(\pi-\gamma,\pi-\xi) \}$. 
To resolve the ambiguities between the strong phase and the weak phase we
can use this method on two or more modes since $\gamma$ will be the same
for each mode while $\xi(k,X)$ should be different.  This is a simple
generalization of the GLW method discussed in section~\ref{section_2}; in
that case, $c(X)=\xba c(X)$ while $\delta_X=n\pi$.

More generally, if we do consider two decay modes with different strong
phases (${\rm mod}~\pi$), we can dispense with the need to know the value
of $b(k)$ as follows:

\bigskip

\bigskip

{\it {\bf\it Case 2:}
For two distinct modes $\{(k,X_1),(k,X_2)\}$
the quantities $d(k,X_i)$ and $\xba d(k,\xba X_i)$ are known. In addition 
$a(k)$ is known  but not $b(k)$. }

\bigskip

In this case we assume that one cannot easily obtain the branching
fraction $b(k)$ using the semileptonic tag or by any other means. However,
if $d$ and $\xba d$ are known for two different modes, we can solve for
the missing information ($\gamma$ and $b(k)$ and $\xi$) 
up to a discrete set of ambiguities.

Specifically, for $D^0$ decay modes $X_1$ and $X_2$ we assume that we have
measured the quantities $\{ a(k)$, $c(X_1)$, $c(X_2)$, $c(\xba X_1)$,
$c(\xba X_2)\}$ as well as $d(k,X_i)$ and $\xba d(k,\xba X_i)$. There are
therefore four unknowns that must be solved for: $\{ b(k)$, $\xi_1$,
$\xi_2$, $\gamma\}$. To do this we use the four equations:

\begin{eqnarray}
d(K,X_1)&=& 
a(K)c(X_1)+b(K)c(\overline X_1)
+2\sqrt{a(K)b(K)c(X_1)c(\overline X_1)}\cos(\xi_{1}+\gamma) 
\nonumber\\
\xba d(K,\xba X_1)&=& 
a(K)c(X_1)+b(K)c(\overline X_1)
+2\sqrt{a(K)b(K)c(X_1)c(\overline X_1)}\cos(\xi_{1}-\gamma) 
\nonumber\\
\label{firstmode}\\
d(K,X_2)&=& 
a(K)c(X_2)+b(K)c(\overline X_2)
+2\sqrt{a(K)b(K)c(X_2)c(\overline X_2)}\cos(\xi_{2}+\gamma) 
\nonumber\\
\xba d(K,\xba X_2)&=& 
a(K)c(X_2)+b(K)c(\overline X_2)
+2\sqrt{a(K)b(K)c(X_2)c(\overline X_2)}\cos(\xi_{2}-\gamma) 
\nonumber\\
\label{eqndd}
\end{eqnarray}

\noindent 
To solve these, 
let us define the quantities.

\begin{eqnarray}
& u_i = {b(k) c(\xba X_i)\over a(k) c(X_i)};  \ \ \ \ 
y_i ={d(k,X_i)-\xba d (k,\xba X_i)\over 2 a(k) c(X_i)}; \ \ \ \
z_i ={d(k,X_i)+\xba d (k,\xba X_i)\over 2 a(k) c(X_i)}-1; & \nonumber\\
& \rho= {c(X_1) c(\xba X_2)\over c(\xba X_1) c(X_2)}={u_2\over u_1};
\ \ \ \ \delta = z_1^2-z_2^2/\rho-2(z_1-z_2)u_1+(1-\rho)u_1^2;  &
\nonumber\\
& \epsilon= y_1^2-y_2^2/\rho;\ \ \ \  
Q=\sin^2\gamma; & 
\label{solvdef}
\end{eqnarray}

\noindent where $y_i$, $z_i$, $\epsilon$ and $\rho$ are known directly
from experiment and $u_i$ and $Q=\sin^2\gamma$ must be solved for.

The equation which $u_1$ must satisfy is easily derived: 

\begin{eqnarray}
4 u_1 \delta\epsilon  = (\epsilon-\delta)(y_1^2\delta-(z_1-u_1)^2\epsilon)
\label{quartic1}
\end{eqnarray}

\noindent Since $\delta$ is second order in $u_1$ this equation is in
general a quartic equation which may have up to 4 real roots. For each
real root $(u_1)_{k}$ (where $k=1,\dots,4$ indexes the solutions of
eqn.~(\ref{quartic1})), $\sin^2\gamma$ is given by:

\begin{eqnarray}
\sin^2\gamma\equiv Q={\epsilon\over \epsilon-\delta}
\end{eqnarray}

\noindent where $\delta$ is given in terms of $u_1$ by
Eq.~(\ref{solvdef}). Each root leads to a four fold ambiguity in the
determination of $\gamma$;  taking up to four roots together there is
therefore $\leq 16$ fold ambiguity in the determination of $\gamma$. To
reduce this ambiguity (especially if it should turn out that all 16
possibilities manifest), it is therefore helpful if observations are made
of at least 3 modes of $D^0$ decay (for a given $k$) in which case only an
overall four fold ambiguity in $\gamma$ remains. Since this method
determines $Q\equiv \sin^2\gamma$ it, therefore, cannot distinguish
between the solutions $\{\pm\gamma$, $\pi\pm\gamma\}$.

For each solution the corresponding total strong phase difference $\xi_i$ 
is then determined without further ambiguity by:

\begin{eqnarray}
\sin\xi_i&=&{-y_i\over 2\sqrt{u_i}\sin\gamma}\nonumber\\
\cos\xi_i&=&{z_i-u_i\over 2\sqrt{u_i}\cos\gamma} \label{xi_expr}
\end{eqnarray}

Another way to resolve the discrete ambiguity is to determine
independently the phase difference: 

\begin{eqnarray}
\Delta\xi &=& \xi_2-\xi_1=\delta_{X_2}-\delta_{X_1}
\end{eqnarray}

\noindent from the study of $D^0$
decays~\cite{ads,charm_X_ads}. It is related to the value of
$\gamma$ and $u_1$ through equation (\ref{xi_expr}) and in general only
two values of $\gamma$ and $u_1$ will give the correct value of
$\Delta\xi$, the remaining two fold ambiguity being between $\gamma$ and
$\gamma+\pi$.

To qualitatively understand the solutions of these equations, it is useful
to consider a plot of $\gamma$ versus $b(k)$. First, let us assume that we
have perfect experimental information.  For a given decay mode $X_i$ where
we know $\{a(k)$, $c(X_i)$, $c(\xba X_i)$, $d(k,X_i)$, $\xba d(k,\xba
X_i)\}$ while $\{\xi_i$, $\gamma$, $b(k)\}$ remain unknown. The two
equations~(\ref{firstmode}) for the mode $X_i$ give a locus of points in
the $\gamma-b(k)$, or equivalently the $\gamma-u_i$ plane when $\xi_i$ is
eliminated. Let us now consider the properties of this curve.

Inspection of these equations shows that they are left unchanged under 
the transformations:

\begin{eqnarray}
(\gamma,\xi_i)&\to&(\pi-\gamma,\pi-\xi)
\nonumber\\
(\gamma,\xi_i)&\to&(-\gamma,-\xi)
\nonumber\\
(\gamma,\xi_i)&\to&(\pi+\gamma,\pi+\xi)
\end{eqnarray}

\noindent Thus it follows that the curve is periodic with respect to
$\gamma\to\gamma+\pi$ and that the curve is also symmetric with respect to
$\gamma\to \pi-\gamma$.  The curve in the range $0\leq\gamma\leq\pi/2$ can
therefore be reflected through $\gamma=0$ and $\gamma=\pi/2$ to get the
entire curve.

For $\gamma=n\pi$, the two cosines in eq.~(\ref{eqndd}) are the same so if
there is any CP violation the equations are inconsistent. This is obvious
from the physics since $\gamma$ is the only CP violating parameter; thus
$\gamma=n\pi$ implies there is no CP violation.  Conversely, if a finite
amount of CP violation is observed, some bound can be placed on $\gamma$. 
In particular, a lower bound $Q_{min}$ can be placed on $Q$:

\begin{eqnarray}
Q>Q_{min}={1\over 2}(1+z_i)(1-\sqrt{1-\alpha^\prime(k,X_i)^2})
\label{gammalimit}
\end{eqnarray}

\noindent
and $\alpha^\prime(k,X_i)$ is the CP asymmetry  defined by:

\begin{eqnarray}
\alpha^\prime(k,X_i)=
{d(k,X_i)-\xba d(k,\xba X_i)
\over d(k,X_i)+\xba d(k,\xba X_i)}
=
{y_i\over 1+z_i}.
\end{eqnarray}

\noindent We also define $\gamma_{min}$ to be the angle in the first
quadrant such that $Q_{min}=\sin^2\gamma_{min}$. Since $\gamma_{min}$
represents the extreme left edge of the figure eight curve, there is a
unique value of $u_i$ which gives $Q=Q_{min}$ we will denote this by $\hat
u_i$ which is given by:

\begin{eqnarray}
\hat u_i
=
z_i+2(1-Q_{min})
\label{uhatdef}
\end{eqnarray}

Likewise, 
there can be no CP violation if $u\to 0$ or $u\to \infty$ and so the 
observation of CP violation implies an upper and lower bound on 
the value of $u$ and hence $b$:

\begin{eqnarray}
u_{max}&=&(1+\sqrt{z_i-|y_i|+1})^2\nonumber\\
u_{min}&=&(1-\sqrt{z_i+|y_i|+1})^2
\label{ulimit}   
\end{eqnarray}

\noindent which leads to the bounds $b_{min} \leq b(k)\leq b_{max}$ where
$b_{min}=a(k)c(X)u_{min}/c(\xba X)$ and $b_{max}=a(k)c(X)u_{max}/c(\xba X)$.

Eliminating $\xi_i$ from the two equations (\ref{firstmode}) gives a 
quadratic equation for $u_i$ 
for a given value of $\gamma$
which defines the set of solutions in the $\gamma-u_i$ plane:

\begin{eqnarray}
Qu_i^2-2(z_i+2(1-Q))Qu_i+(Qz_i^2+(1-Q)y_i^2)=0.
\label{drawcurve}
\end{eqnarray}

This is quadratic in $u_i$;  therefore, there are zero, one or two
solutions for $u_i$ at any given value of $\gamma$.  In particular, if
$Q<Q_{min}$ there are no solutions while for $Q=Q_{min}$ there must be
exactly one solution (i.e. $u=\hat u$).  It is also the case that when
$Q=1$, for instance, when $\gamma=\pi/2$, there is again exactly one
solution $u_i=z_i$. This follows from the fact that in the sum $d+\xba d$
the interference term vanishes and a value of $u_i$ may thus be obtained. 
Taking the $\gamma$ axis horizontal, the curve in the range
$0\leq\gamma\leq\pi$ therefore has the topology of a lazy eight centered
about the vertical line $\gamma=\pi/2$ which crosses itself at
$(\pi/2,z_i)$.

In Fig.~1 we show a plot of $u_i$ versus $\gamma$ for $z_i=1.5$ and
$y_i=0$, $1$, $2$. The bounds given by
Eqns.~(\ref{gammalimit},\ref{ulimit}) are indicated with the rectangular
boxes. Clearly the greater the value of $y_i$ (and the greater the amount
of CP violation), the greater the bounds which may be placed on $u_i$ and
$\gamma$ through these inequalities by considering just one mode.  Thus
single modes which have a high degree of CP violation are quite desirable
since they lead to the most restrictive bounds in the $\gamma-u_i$ plot.
Indeed, as we have argued above, larger CP violation is more likely to
arise in the case of non-CP eigenstate final states (such as $\xba D\to
K^+\pi^-$) as compared to CP-eigenstate modes.

Let us now turn our attention to the case when we have two modes present. 
If we had perfect data concerning each mode, we would generate two of
these lazy-eight curves in the $\gamma-b(k)$ plane.  In general, these
curves will therefore intersect in as many as four points in the range
$0\leq\gamma\leq\pi/2$ which correspond to the solution of
eq.~(\ref{quartic1}). If the data was inconsistent, then the curves would
miss each other corresponding to a situation where eq.~(\ref{quartic1})
has no real solutions for $\gamma$.  In order to understand how this might
play out, let us now consider a scenario for the not yet observed
branching ratios.

\section{Attainable Accuracy}\label{section_5}

\subsection{Estimates of Branching Ratios}\label{section_5_1}

For the purpose of illustrating the ability to extract $\gamma$ by
combining measurement of several modes, we need to make a simple estimate
of the branching ratios of DCS decay modes of $D$  which have not yet been
measured.  We will proceed by relating these modes to the well measured
Cabibbo-allowed ones and then use factorization to break down $D\to
M_1M_2$ to $\langle M_2|J_{cd} |D\rangle \langle M_1|J_{us}|D\rangle$. 
The SU(3) breaking of the decay constants is used in keeping track of the
piece $\langle M_1|J_{us} | 0\rangle$. Single-pole dominance allows one to
keep track of the SU(3) breaking in $\langle M_1|J_{cd}|D\rangle$. In
the final estimated rates, we also factor in the small difference in phase
space.

For a concrete example let us consider the DCS mode $D^0\to K^+\pi^-$
which we relate in this procedure to CBA counterpart $D^0\to K^-\pi^+$. 
Using this reasoning the amplitude will be proportional to: 

\beq
A(D^0\to K^+\pi^-) \propto  \lambda^2
f_K (m^2_D-m^2_\pi) 
\left(
1-\frac{m^2_K}{m^2_{D^\ast}}
\right)^{-1} 
\label{eqbtwo}
\eeq

\noindent Thus

\beq
\left |
\frac{A(D^0 \to K^+\pi^-)}{A(D^0 \to \pi^+K^-)} 
\right |
= 
\lambda^2
\left(\frac{f_K}{f_\pi} \right) \left(\frac{1 - m^2_\pi/m^2_D}{1 -
m^2_K/m^2_D} \right) 
\left(\frac{1 - m^2_\pi/m^2_{D^\ast_s}}{1 - m^2_K /m^2_{D^\ast}} \right) 
\label{eqbthree} \eeq

\noindent In this instance there is no phase space correction. Using
$\lambda=0.22$, $f_K=160$ MeV, $f_\pi=132$ MeV and other masses we thus
get,

\beqa
\frac{BR(\xba D^0\to K^+\pi^-)}{BR(D^0 \to \pi^+K^-)} & = & 1.88
\lambda^4 \nonumber \\
& = & 5.3\times 10^{-3} \label{eqbfour} 
\eeqa

\noindent Therefore $BR(D^0\to \pi^+K^-)=3.83 \times10^{-2}$~\cite{pdb} 
gives:

\beq 
BR (D^0 \to K^+\pi^-) = (2.0\pm .7) \times 10^{-4} \label{eqbfive}
\eeq

\noindent which should be compared to the measured value~\cite{pdb}

\beq 
BR(D^0\to K^+\pi^-) = (2.8\pm 0.9) \times 10^{-4} \label{eqbsix}
\eeq

\noindent We will normalize our predictions for all the DCS modes to
the measured central value.

For the other DCS modes of interest to us we can now use relations
amongst DCS modes and their CBA counterpart.  Thus, e.g., for 
$D^0\to K^+\rho^-$ we should have:

\beq
\frac{A(D^0\to K^+\rho^-)}{A(D^0\to K^+\pi^-)} 
\approx 
\frac{A(D^0\to
\pi^+ K^{\ast -})}{A(D^0 \to \pi^+K^-)} \label{eqbseven}
\eeq

Similar scaling relations for the other two modes are:

\beqa
\frac{A(D^0\to K^+a^-_1)}{A(D^0\to K^+\pi^-)} 
&\approx & 
\frac{A(D^0\to \pi^+ K^-_1 (1270))}{A(D^0 \to\pi^+ K^-)} 
\nonumber\\
\frac{A(D^0\to K^{\ast+}\pi^-)}{A(D^0 \to K^+\pi^-)} 
& \approx &
\frac{A(D^0\to\rho^+K^-)}{A( D^0\to\pi^+K^-)} 
\label{eqbnine}
\eeqa

\noindent 
The resulting branching ratio
for all the DCS modes of interest are given in Table~\ref{tabone}.

To complete our sample calculation, we need to estimate
the expected magnitudes of $\{a,b\}$

Starting with $a(k)$, we will extrapolate from the observed branching
fraction for related $B$ decays\cite{pdb}:

\begin{eqnarray}
Br(B^-\to \pi^- D^0)&=&(5.3\pm 0.5)\times 10^{-3} \nonumber\\
Br(B^-\to \rho^- D^0)&=&(13.4\pm 1.8)\times 10^{-3}.
\end{eqnarray}

\noindent
Multiplying this by 
$\sin^2\theta_C$ one obtains the estimates for $a(k)$:

\begin{eqnarray}
a(K)&\approx&2.6\times 10^{-4} \nonumber\\
a(K^*)&\approx&6.6\times 10^{-4} \label{aest}
\end{eqnarray}

The estimation of $b(k)$ is more uncertain since it is a color suppressed
process.  Thus, to estimate the branching ratio for $B^-\to k^- \xba D^0$
we use the fact that the quark level diagram for this process is color
suppressed with respect to $B^-\to k^- D^0$ and take the color suppression
to be simply a factor of $1/N_c$. Folding in all the appropriate CKM
elements an estimate for the branching ratio may be obtained from the
previous estimate of $a(k)$ (taking $N_c=3$):

\begin{eqnarray}
{b(k)\over a(k)} \approx \left[   {|V_{ub}||V_{cs}|\over N_c |V_{cb}|
|V_{us}|} \right]^2\approx .015 \label{colorrat}
\end{eqnarray}

\noindent
and so for the two specific cases:

\begin{eqnarray}
b(K)&\approx&  4.0 \times 10^{-6} \nonumber\\
b(K^*) &\approx& 10.0  \times 10^{-6} \label{best}
\end{eqnarray}

\subsection{CP Violation in Single Modes}\label{section_5_2}

Let us now perform some sample calculations using the above estimated
branching ratios in order to illustrate what precision might be obtained
by this method.

First, we will consider what can be learned through the use of data in a
single mode. As discussed above, this will not allow one to obtain an
exact value of $\gamma$ but can give a bound on $\gamma$ if large CP
violation is present. We will therefore consider how many $B$ decays are
needed to have a measurable signal of CP violation and what bounds on
$\gamma$ would follow.  We will next consider the precision for extracting
$\gamma$ by combining the data from several modes.

To illustrate this let us consider the specific case of CP violation when
$X=K^+\pi^-$.  The observed~\cite{pdb} branching ratios for the $D$ decays
are:

\begin{eqnarray}
c(K^+\pi^-)&=&(2.9\pm 1.4)\times 10^{-4}\nonumber\\
c(K^-\pi^+)&=&(3.83\pm 0.12)\times 10^{-2}
\end{eqnarray}

\noindent Where the partial partial rate asymmetry is given by:

\begin{eqnarray}
\alpha^\prime(k,X)= { d(k,X)-\xba d(k,\xba X)\over d(k,X)+\xba d(k,\xba X) }
\end{eqnarray}

\noindent
which from equation (\ref{eqnd}) can be written:

\begin{eqnarray}
\alpha^\prime(k,X)= - { R(k,X) \sin(\zeta_k+\delta_X)\sin\gamma \over
1+R(k,X)\cos(\zeta_k+\delta_X)\cos\gamma } \label{praeqn}
\end{eqnarray}

Let $N_B$ be the total number of $B^\pm$. In a given mode, let $E_i$ be 
the acceptance times efficiency of a detector and let us denote:

\begin{eqnarray}
\tilde N_B
=
E_i N_B
\end{eqnarray}

Thus if $\nup(k,X)$ is the number of charged $B^\pm$ required to observe
the partial rate asymmetry at a $3\sigma$ level with an ideal detector, it
is related to the actual number, $N_B^{3\sigma}(k,X)$, of $B^\pm$ by $E_i
N_B^{3\sigma}(k,X)=\nup(k,X)$. In terms of the CP asymmetry, $\nup$ is
given by:

\begin{eqnarray}
\nup(k,X)={18\over [\alpha^\prime(k,X)]^2 [d(k,X)+\xba d(k,\xba X)]}
\label{upseqn}
\end{eqnarray}

In order to obtain a large value of $\alpha^\prime$ it is clearly
necessary to have a large value of $R(k,X)$ since $|\alpha^\prime|\leq
R(k,X)$. As defined, $0\leq R(k,X) \leq 1$ where $R(k,X)$ is maximized
when $a(k)c(X)\approx b(k)c(\xba X)$.  It should be clear that this will
happen if the two channels have roughly the same amplitude. In particular,
using the numbers in the estimates above, $R(K^-,K^+\pi^-) \approx
R(K^{*-},K^+\pi^-) \approx 0.94$. On the other hand if we had considered
the case where $D^0$ decays in a CBA mode, then we would have obtained a
much smaller value, $R(K^-,K^-\pi^+) \approx R(K^{*-},K^-\pi^+) \approx
.02$ and so CP violation would be small.

To completely specify $\alpha^\prime$ and $\nup$, of course we also need
to know $\cos\xi\cos\gamma$ and $\sin\xi\sin\gamma$.  Since these are
totally unknown, to get a rough idea of the experimental
requirements we will take the sample values $\cos(\xi)\cos\gamma=0$ and
$\sin(\xi)\sin\gamma=1/2$.

In this example we obtain

\begin{eqnarray}
\alpha^\prime(K^-,K^+\pi^-) \approx \alpha^\prime(K^{*-},K^+\pi^-)
&\approx& 0.47 \nonumber\\
\nup(K^-,K^+\pi^-)&\approx& 17.6\times 10^7  \nonumber\\
\nup(K^{*-},K^+\pi^-)&\approx& 7.0\times 10^7 
\end{eqnarray}

\noindent
Using the estimated branching ratios in Table~1, we can perform a similar 
estimate for some of the other possible modes:

\begin{eqnarray}
\nup(K^-,K^+\rho^-) &\approx&  6.3\times 10^7  \nonumber\\
\nup(K^{*-},K^+\rho^-) &\approx&  2.5\times 10^7  \nonumber\\
\nup(K^-,K^+a_1^-) &\approx&  9.3\times 10^7 \nonumber\\
\nup(K^{*-},K^+a_1^-) &\approx&  3.7\times 10^7 \nonumber\\
\nup(K^-,K^{*+}\pi^-) &\approx&  13.6\times 10^7 \nonumber\\
\nup(K^{*-},K^{*+}\pi^-) &\approx&  5.4\times 10^7 
\end{eqnarray}

\noindent where the asymmetries $\alpha^\prime$ for these modes are given
in Table~2.

Since each of these modes as well as several other possibilities have a
different values of $\xi_X$ it is at least likely that a few 
instances of this kind
of CP violation can be observed in the $\nup\sim 10^8$ range.

For comparison let us consider one of the CP eigenstate modes as in GLW,
using the above numbers. In this case $X=X_{CP}$ is a CP eigenstate. In
particular, let us take the mode $X_{CP}=K_S\pi^0$.  Using the $k=K^*$
case we have as before $a(K^*)=6.6\times 10^{-4}$; $b(K^*)=10.0\times
10^{-6}$.  In this case $c(K_s\pi^0)=\xba c(K_s\pi^0)=1.05\times 10^{-2}$.
From this we get $R=0.24$. Here $\delta=0$ so making a similar assumption
as above concerning the phases, that $\sin\zeta\sin\gamma\approx 1/2$ and
$\cos\zeta\cos\gamma\approx 0$ we obtain
$\alpha^\prime(K^*,K_S\pi^0)\approx 0.12$. Substituting this into equation
(\ref{upseqn}) we obtain $\nup\approx 9.0\times 10^{7}$.

In principle we can improve this by folding in information from all possible 
modes that are $CP$ eigenstates\footnote{In particular we add the 
$d(k,X)$ for CP even states to $\xba d(k,X)$ for CP odd states and vice 
versa}. Some particular $D^0$ decay  modes which are CP eigenstates are

\begin{eqnarray}
\{ 3K_s,~ K_s\eta,~ K_s\rho^0,~ K_s\omega,~ K_s\eta^\prime,~ K_sf_0, K_s\phi,
\nonumber\\
K_s f_2,~ \pi^+\pi^-,~  K^+K^- \}
\end{eqnarray}

\noindent Using the observed branching fraction in \cite{pdb} these have a
total branching fraction of about $5\%$ (if the corresponding modes with
$K_L$ are included, this total roughly doubles). Using the effective $c$
obtained with the CP non-eigenstate method from all of these modes
together, we get $\nup(K^{*-},X_{CP})\approx 1.9\times 10^{7}$ which is
roughly what we obtained in the single mode $\nup(K^{*-},K^+\rho^-)\approx
2.5\times 10^{7}$. Thus, although the event rate for the various DCS modes
is smaller than the CP eigenstate modes, the CP violating asymmetries are
larger and $\nup$ is comparable. The latter case does have the following
potential advantages

\begin{enumerate} 

\item Each mode has a different strong phase difference and so it is
perhaps more likely that one will have a total strong phase near
$\pm\pi/2$ (i.e. maximal CP violation). 

\item Since $\alpha^\prime$ tends to be larger, the bounds placed on
$\gamma$ will tend to be more restrictive. 

\end{enumerate}

In order to illustrate these two points, Fig.~2 shows the locus of allowed
points on a plot of $\gamma$ versus $b(K^*)$ which would be obtained with
the example above in the case of a $K^+\pi^-$ final state (solid line) and
a $K_S\pi^0$ final state (dashed line) assuming an ideal measurement.
Clearly, in this instance the restriction on $b(k)$ and $\gamma$ is much
tighter for the $K^+\pi^-$ final state than for the $K_s\pi^0$ final
state.

\subsection{Combining Information From Several Modes}\label{section_5_3}

Let us now consider the degree of accuracy in the determination of
$\gamma$ that may be obtained through the combination of various modes. To
do this, we will do an illustrative calculation of the 90\% confidence
levels one can obtain on the $\gamma-b(k)$ plane with $\nup=10^8$.

To produce a scenario, we will consider some specific values of $\gamma$
and $\xi_i$ together with the branching ratios above and then calculate
the values of $d$ and $\xba d$ which would be relevant. If we now take
these as being the experimental value we can then consider the likelihood
as a function of $\gamma$ and $b(k)$.

The modes we will consider are ($D^0\to K^+\pi^-$, $D^0\to K_S\pi^0$,
$D^0\to K^+\rho^-$, $D^0\to K^+a_1^-$, $D^0\to K^0\rho^0$ and $D^0\to
K^{*+}\pi^-$).  We will take $\tilde N_B=10^8$ and consider $k=K^*$.  It is
important to note that in this example we consider statistical errors
only. It is clear that in order to perform such a study, 
systematic errors must be under control.

The exact experimental results will, of course depend on the strong 
phases and $\gamma$. For the purposes of illustration, let us assume that 
$\gamma=60^\circ$ and the strong phases:

\begin{eqnarray}
\xi(K^+\pi)=10^\circ  & \xi(K_s\pi^0)=20^\circ \nonumber\\
\xi(K^+\rho^-)=30^\circ  & \xi(K^+a_1^-)=40^\circ \nonumber\\
\xi(K_s\rho^0)=200^\circ  & \xi(K^{*+}\pi^-)=50^\circ 
\end{eqnarray}

\nonumber Numerically, the values of $d$ and $\xba d$ are given in Table~2
where the $d_i$, $\xba d_i$ entries are given in units of $10^{-8}$.

In Fig~3a and 3b we show the likelihood contours as a function of
$\gamma$ and $b(K^*)$.  In Fig~3a we consider only the data from the
$K^+\pi^-$ and $K_s\pi^0$ final states\footnote{Note that it has been
suggested\cite{ads_cross_glw} that the combination of a DCS mode and a CP
eigenstate mode (as in this example) offers some advantage because of the
larger statistics that can be obtained by combining different eigenstate
modes. This may be the case although how well a particular combination
works also depends to a great degree on the strong phases.}.  The solid
curve shows the locus of solutions which explain the $K^+\pi^-$ data
while the dashed curve shows the solutions which explain the $K_s\pi^0$
data. As can be seen there are four intersections which is the case in
general when just two modes are considered. The contour regions
correspond to 68\% and 90\% confidence levels based on $\nup=10^8$. In
Fig.~3b we consider the situation where data from all six modes is used. 
In this figure, the solution for $K^+\pi^-$ is shown in solid, $K_S\pi^0$
with short dashes, $K^+\rho^-$with long dashes, $K^+a_1^-$ with the
dash-dot line, $K_S\rho^0$ with the dash-dot-dot line and $K^{*+}\pi^-$
with the dash-dash-dot line.  As can be seen from this figure, the
correct value of $\gamma$ and $b(k)$ is now identified with the only
ambiguity being the four fold ambiguity in $\gamma$ between $\pm\gamma$
and $\pi\pm\gamma$.

In Fig.~4 we have projected the likelihood from Fig.~3b onto the $\gamma$
axis where we have considered the case of $\gamma=15^\circ$, $30^\circ$,
$60^\circ$ and $90^\circ$ which are indicated by the curves peaked at
those values of $\gamma$. In each of these cases, the 90\% confidence
interval is about $\pm 10^\circ$ about the solution.

It should be realized that three body states $K^+ \rho^-$, $K_s \rho^0$
and $K^{*+} \pi^-$ can all lead to the common final state $K_s
\pi^+\pi^-$. If one examines the distribution in phase space, then the
vector resonances overlap to some extent and the channels will interfere
with each other. In the following section, we will discuss how the
additional information implicit in this situation can assist in extracting
the value of $\gamma$.

\section{Using Three Body Decays}\label{section_6}

Here we will consider the generalizations of the two approaches considered
in section~\ref{section_4} to the case of a three body decay.  First of
all, we can consider the three body decay as consisting of a number of
quasi two body channels which we can regard as distinct modes and find a
solution for $b(k)$ and $\gamma$. A second approach is to regard each
point of the Dalitz plot as a distinct mode. We can then apply the
inequalities eqns.~(\ref{gammalimit},\ref{ulimit}) at each point. Since
all of these inequalities must be true simultaneously, a very stringent
bound can generally be placed on $\gamma$ and $b(k)$. In fact we will
argue that for at least some points this inequality is an equality so the
limit given by such an argument should in fact give $\gamma$ and $b(k)$.

As an example we will consider in particular the case of $D^0$, $\xba
D^0\to K^+\pi^-\pi^0$.  In this case the CBA decay $\xba D^0\to
K^+\pi^-\pi^0$ has been experimentally studied by the E687
collaboration~\cite{e687ref}. The data they obtain is fit to an amplitude
to a general multi-channel 3-body decay form:

\begin{eqnarray}
{\cal M}(\xba D^0\to K^+\pi^-\pi^0)
= a_0e^{i\delta_0} + \sum_i a_i exp(i\delta_i) B(a,b,c|r)
\end{eqnarray}

\noindent
where $r$ is a label for the resonance and
$a$,$b$ and $c$ are labels for the three final state particles which are
permuted so that $(a,b)$ forms the resonance that a given term represents.
Thus, the function $B$ is given by 
$B(a,b,c|r)=BW(a,b|r){\cal S}(a,c)$ where if $J_r$ is the spin of $r$,

\begin{eqnarray}
BW(a,b|r)&=&-(s_{ab}-m_r^2+i\Gamma_r(s_{ab}) m_r)^{-1} \nonumber\\
&& \nonumber\\ {\cal S}(a,c)&=&\left\{ \begin{array}{ccc}
1                                      &{\rm if}&  J_r=0  \\
-2 \vec P_a \cdot \vec P_c             &{\rm if}&  J_r=1  \\
2  \left(3 \left(\vec P_a\cdot\vec P_c\right)^2 -|\vec P_a|^2|\vec 
P_c|^2\right)    
&{\rm if}&  J_r=2  \\ \end{array} \right.
\end{eqnarray}

\noindent
where $s_{ab}=(p_a+p_b)^2$ and  $\vec P_a$ and $\vec P_c$ are the 3-momenta
of $a$ and $c$ {\it in the $ab$ rest frame} and 

\begin{eqnarray}
\Gamma_r(s_{ab})
=
\Gamma_r(m_r^2)
\left (
m_r^2 \left |\vec P_a\right |^2
\over
(m_r^4+m_a^4+m_b^4-2m_r^2m_a^2-2m_r^2m_b^2-2m_a^2m_b^2)
\right )^{J_r+{1\over 2}}
\end{eqnarray}

\noindent
Here $m_r$ is the mass of the resonance and
$\Gamma_r$ is the energy dependent width.

Thus, the general model used in~\cite{e687ref} for a 3-body decay includes
a number of resonance channels, each of which may have a different spin
and phase $\delta_i$ together with a non-resonant term (i.e.\
$a_0\exp(i\delta_0)$)

Specifically, for $\xba D^0\to K^+\pi^-\pi^0$ they use a non-resonant 
term together with three channels:

\begin{eqnarray}
\xba{\cal M} &=&  
a_0          \exp({i\delta_0}) 
+a(K^{*0})   \exp({i\delta_{K^{*0}}})    B(K^+\pi^-\pi^0|K^{*0}) 
\nonumber\\
&&
+a(K^{*+})   \exp({i\delta_{K^{*+}}})    B(K^+\pi^0\pi^-|K^{*+}) 
\nonumber\\
&&
+a(\rho-)    \exp({i\delta_{\rho^{-}}})  B(\pi^-\pi^0K^+|\rho^-) 
\label{cbardef}
\end{eqnarray}

\noindent In this analysis\cite{e687ref} the decay 
fractions\footnote{
The decay fraction for a given channel $X$ is the rate of
$D^0\to K^+\pi^-\pi^0$ if all channels aside from $X$ are turned off
divided by the total rate of $D^0\to K^+\pi^-\pi^0$ with all channels
present. Thus, for instance, the decay fraction through $K^{*+}\pi^-$ is 
$Br(\xba D^0 K^{*+}\pi^-)Br(K^{*+}\to K^+\pi^0)$. Due to interference 
effects, the decay fractions need not add up to 1.
}
for each
channel and the relative phases were determined which we have summarized
in Table~3.  In our calculation we will use these decay fractions
together with the global average for the total branching ratio, $Br(\xba
D^0\to K^+ \pi^0\pi^0) = 13.9\pm0.9\%$ given in~\cite{pdb}. For the
purposes of our illustrative calculations we will assume that these
quantities take on their central values.

The DCS decay $D^0\to K^+\pi^-\pi^0$ has not likewise been investigated
experimentally, so for the purposes of making numerical estimates, we will
use a model based on SU(3).  First we write a decomposition of ${\cal M}$ 
similar to that of  eq.~(\ref{cbardef}):

\begin{eqnarray}
{\cal M}( D^0\to K^+\pi^-\pi^0)
&=&  
a_0^\prime          \exp({i\delta_0^\prime}) 
+a^\prime(K^{*0})   \exp({i\delta_{K^{*0}}^\prime})   B(K^+\pi^-\pi^0|K^{*0}) 
\nonumber\\
&&
+a^\prime(K^{*+})   \exp({i\delta_{K^{*+}}^\prime})   B(K^+\pi^0\pi^-|K^{*+}) 
\nonumber\\
&&
+a^\prime(\rho^{-}) \exp({i\delta_{\rho^{-}}^\prime}) B(\pi^-\pi^0K^+|\rho^-)
\label{cdef} 
\end{eqnarray}

\noindent
Using the relations between the branching  
ratios to two body decays discussed 
in section~\ref{section_5_1}, we can relate the magnitudes 
$|a^\prime (K^{*0})|$,
$|a^\prime (K^{*+})|$ 
and
$|a^\prime (\rho^-)|$
to 
$|a   (K^{*0})|$,
$|a   (\rho^{-})|$ 
and
$|a   (K^{*+})|$
by regarding them as amplitudes for quasi-two body decays.
In addition we set $|a_0^\prime|=\lambda^2|a_0|$. 
Assuming exact SU(3) for  
the phases of the DCS channels, they will be:

\begin{eqnarray}
\delta_0^\prime        = \delta_0                   \ \ \ 
\delta_{K^{*0}  }^\prime  = \pi+\delta_{K^{*0}}      \ \ \ 
\delta_{K^{*+}  }^\prime  = \delta_{\rho^{-}}         \ \ \ 
\delta_{\rho^{-}}^\prime  = \delta_{K^{*+}}            
\end{eqnarray}

This model then predicts:

\begin{eqnarray}
Br(D^0\to K^+\pi^-\pi^0) = 7.8\times 10^{-4} 
\end{eqnarray}

The only free parameters now are the strong phase difference $\xi$ between
$B^-\to k^- D^0 $ and $B^- \to k^- \xba D^0$, the overall branching ratios
$a(k)$ and $b(k)$ and of course $\gamma$. If we use the estimates of
$a(K^*)$ and $b(K^*)$ discussed in section~\ref{section_5}, we can
therefore obtain the Dalitz plots of the $K^+\pi^-\pi^0$ final state which
resulting from interference of $D^0$ and $\xba D^0$ channels.

Fig.~5a shows such a plot with $\gamma=90^\circ$ and $\xi=70^\circ$ which
is intended to represent a sample of $\nup=10^{10}$.  The upper plot
represents $B^-$ decays where the variables used are
$s=(p_{\pi-}+p_{K+})^2$ and $t=(p_{\pi0}+p_{pi-})^2$ while the lower plot
represents the CP conjugate.

It is clear that in this case there is significant amount of CP
violation. In Fig.~5b we have taken $\xi=160^\circ$ and again it is clear
that CP violation is present but here it is more subtle. There are about
the same number of points in each of the plots but the distribution
changes going from $B^-$ to $B^+$.

In these plots, the three resonant bands are prominent, one could isolate
the resonant contributions by fitting the plot to a model such as
discussed above and thus obtain four modes (the three resonances plus the
non-resonant contributions) to feed into the analysis of the previous
section.  Obviously there is more information implicit in these
distributions. For instance, one would have additional constraints since
one would know the relative strong phase difference between each of the
$D$ decay modes. 

In particular, for this model one has the magnitudes of four amplitudes
each for $B^-$ and $B^+$ decay (giving 8 parameters) and three phase
differences\footnote{one cannot determine an overall phase so only the
differences between the $\delta$'s can be extracted from measurements}
giving a total of 14 parameters. The three unknowns in this case are just
$\gamma$, $b(k)$ and $\zeta$ so they are well over-determined.

In order to get an idea of how many B mesons are required to extract useful
information from this kind of data, let us consider what is required to
find a 3-$\sigma$ signal of CP violation for this system.  In Fig.~6a we
show a plot of the overall partial rate asymmetry for this mode for
$\gamma=90^\circ$ as a function of the overall strong phase difference
$\eta$. In Fig.~6b we show with the solid line the number of $B^0$
($\nup$) mesons required to give a statistical 3-$\sigma$ signal
for the partial rate asymmetry.

When the partial rate asymmetry is small, for instance in the case where
$\eta=160$ in Fig.~5b, clearly $\nup$ becomes large but of course, we are
not using the information contained in the full distribution. Using the
methods of~\cite{optobsref} if we assume the CBA and DCS 
decays of the $D^0$
are 
understood we can construct an observable or system of weights of
various regions of the Dalitz plot which is optimally sensitive to CP
violation. Using this method, $\nup$ is shown in Fig.~6b with the dashed
curve;  depending on $\eta$, $\nup$ varies between $(0.35-2)\times
10^{8}$.

Another approach to obtaining $\gamma$ is to use the generalization of
the bound given in eqn.~(\ref{gammalimit}). If one fits the Dalitz plots
to continuous curves, then, regarding each point as a separate mode, we
would have $Q_{min}\leq Q$ as a function of the Dalitz plot variables.
Since each lower bound must be valid, the maximum lower bound, $\hat Q =
max(Q_{min})$ will provide the most stringent lower bound on $Q$ so
therefore $Q\geq \hat Q$.

In fact it is not unreasonable to expect that
$\hat Q=Q$. To see this note that
eqn.~(\ref{uhatdef}) tells us that if $Q=Q_{min}$,
then

\begin{eqnarray}
u=z_i+2\cos^2\gamma
=u+2\sqrt{u}\cos\gamma\cos\xi
+2\cos^2\gamma
\end{eqnarray}

so therefore

\begin{eqnarray}
\cos\xi=-\cos\gamma/\sqrt{u}
\label{goodplot}
\end{eqnarray}

\noindent If eqn.~(\ref{goodplot}) is true anywhere on the Dalitz plot,
then $\hat Q=Q$ and one would expect that this condition would apply on
some curve on the dalitz plot. This is illustrated in Fig.~7a where the
locus of points where $Q=Q_{min}$ is shown for $\gamma=60^\circ$ with
$\zeta=0^\circ$ (solid line); $\zeta=30^\circ$ (dashed line);
$\zeta=60^\circ$ (dotted line) and $\zeta=90^\circ$ (dot-dashed line).

To extract $Q$ using this method, it is useful to consider the function

\begin{eqnarray}
f(q)={
\int_{Dalitz} \theta(Q_{min}(s,t)-q) ~ds ~dt 
\over
\int_{Dalitz}  ~ds ~dt 
}
\end{eqnarray}

\noindent where the integral is over the allowed region of the Dalitz plot
and $\theta(x)=0$ if $x<0$ or $1$ if $x\geq 0$.  Thus $f(q)$ is the
fraction of the Dalitz plot such that $Q_{min}\geq q$. 
For values of $q
\to Q$ from below, $f(q)\propto \sqrt{Q-q}$ and so $f^2(q)$ will linearly
extrapolate its endpoint at $q=Q$.

In Fig.~7b we show a plot of $f(r Q)$ where $r=0.7$ and $0.9$ as a 
function of $\gamma$ for 
$\zeta=0^\circ$ (solid);
$\zeta=30^\circ$ (dashed);
$\zeta=60^\circ$ (dotted) and
$\zeta=90^\circ$ (dot-dashed).
From this is apparent that for almost every combination of strong and 
weak phases roughly $20\%$ of the dalitz plot has $Q_{min}\geq 0.9 Q$.

In Fig.~7c we show a graph of $f^2(q)$ for $\zeta=90^\circ$ with $Q=0.25$,
$Q=0.5$, $0.75$.  As can be seen, in all cases a significant fraction of
the Dalitz plot has $Q_{min}$ close to $Q$ and the curves extrapolate
linearly to an endpoint at $q=Q$.

Likewise we can extract $b(k)$ by applying eqn.~(\ref{ulimit}) to each
point of the Dalitz plot. Here one has both an upper and a lower bound on
$b(k)$ so that if one considers the functions:

\begin{eqnarray}
g_{min}(b)&=&
{
\int_{Dalitz} \theta(b_{min}(s,t)-b) ~ds ~dt
\over
\int_{Dalitz}  ~ds ~dt
}
\nonumber\\
g_{max}(b)&=&
{
\int_{Dalitz} \theta(b-b_{min}(s,t)) ~ds ~dt 
\over
\int_{Dalitz}  ~ds ~dt
}
\end{eqnarray}

\noindent the support of $g_{min}$ must lie entirely below $b(k)$
while the support of $g_{max}$ must lie entirely above $b(k)$. As with the
bound in $Q$, the end point of the support in both cases is $b(k)$.

In Fig.~8 we show a graph of $g_{min}^2(b)$ and $g_{max}^2(b)$
as a function of $b/b(k)$ in the case of $D^0\to K^+\pi^-\pi^0$. Here we
take $\gamma=60^\circ$ and $\zeta=90^\circ$. Again a significant portion
of the Dalitz plot can be seen to give values of $b_{min}$ and $b_{max}$
close to the endpoint. 

If $b(k)$ were determined in this way, the information could easily be
combined with the branching ratio to a two body final state (e.g. 
$K^+\pi^-$ or CP eigenstates) to obtain $\gamma$ as discussed in Case~1 of
section~\ref{section_4}.

We can also apply the methods discussed in this section 
to the case where the $D^0$ decays to a 2-body final state (eg.
$D^0\to K^+\pi^-$ or $K_S\pi^0$) but the parent $B^-$ decays to a 3-body
final state, for instance $B^-\to D^0 K_S\pi^-$. In this case, of course
the Dalitz plot variables are those of the parent $B^-$ decay but other
than that, one could solve for $\gamma$ by fitting the the distribution to
a series of amplitudes which could be treated as different modes or by
finding $\hat Q_{min}$ as described above.

\section{Summary and Conclusion}\label{section_7}

In conclusion, we have considered a number of ways of observing the
interference between $B^-\to K^- D^0$ and $B^-\to K^-\xba D^0$ as a means
of cleanly extracting $\gamma$.  In general, this is achieved by
observing final states common to $D^0$ and $\xba D^0$ decay. To fully
determine $\gamma$ with such observations, several different decay modes
must be observed or else the difficult to measure decay rate for $B^-\to
K^-\xba D^0$ must be independently known. 

Thus if for one $D^0$ decay mode $X$ the branching ratios for $B^-\to
K^-X$ and $B^+\to K^+\xba X$ are known, where the $X$ results from the
interference of $D^0$ and $\xba D^0$ channels and if one also knows
$B^-\to K^-\xba D^0$ then $\gamma$ can be determined up to an 8 fold
ambiguity.  The special case where $X$ is a CP eigenstate was originally
considered in \cite{gronauwyler}. 

On the other hand if the branching ratio for $B^-\to K^-\xba D^0$
is not known but there is a large degree of CP violation in the mode, 
then it may be possible to put a lower bound on $\sin^2\gamma$. A 
particularly promising class of decays which could give large CP 
asymmetries are cases where $D^0\to X$ is a DCS decay. In such cases the 
enhancement of the decay $\xba D^0\to X$ which is CBA is balanced by the 
enhanced production amplitude $B^-\to K^- D^0$ which is CLA. The two 
channels thus have roughly equal amplitudes therefore CP asymmetries can 
be large.

If two or more modes $X_1,\dots,X_n$ are measured, then we no longer need
to know $B^-\to K^-\xba D^0$, it can be fit for along with $\gamma$. In
the case of 2 modes there is potentially a 16 fold ambiguity due to the
need for solving of a quartic equation in $\sin^2\gamma$. If 3 or more
modes are considered, then there is only a 4-fold ambiguity since the
system of equations for $\sin^2\gamma$ is over determined.  In our sample
calculation it was found that for $\hat N_B=10^8$, that the 90\% bound on
$\gamma$ found from combining several modes is roughly $15^\circ$ which
is typical although the actual precision depends on the values of the
strong phase differences. 

This approach can be generalized to the case where the $D^0$ undergoes a
3-body (or indeed $n$-body) decay.  If one considers each point in the
dalitz plot to be a single mode, one can obtain a lower bound for
$\sin^2\gamma$ at each point.  In general, one expects that the maximum
lower bound is in fact equal to $\sin^2\gamma$ so this method actually
gives $\gamma$ up to a four fold ambiguity.  The same method may also be
applied to determine $b(k)$ by obtaining upper and lower bounds on $b(k)$
at each point. Alternatively, fitting the Dalitz distributions to a
resonant channel model provides enough information to obtain $\gamma$. 

Although throughout we have assumed that $D\xba D$ mixing is negligible,
in the appendix we show how the effects of such mixing may be eliminated
by using information about the time between the $B^-$ and $D^0$ decays at
the expense of increasing the statistical errors by about $\sqrt{2}$.

\bigskip
\bigskip

This research was supported in part by the U.S. DOE contracts
DE-FG02-94ER40817 (ISU) and DE-AC-76CH00016 (BNL).

\section*{\bf Appendix: The Implications of $D^0 -\xba D^0$ 
Mixing}

In the discussion so far we have explicitly assumed that $D^0\xba D^0$
was negligible. In particular, since we often take advantage of
interferences involving DCS decays which are $O(1\%)$ of the interfering
CBA decay, the total probability of mixing must be less than
$O(1\%)$ for the above formulation to remain valid. 
In this section we consider the generalization to the case where 
$D\xba D$ mixing may be present.

We will argue that for final states which involve the DCS decay of the 
$D^0$ (such as $B^-\to K^- [D^0\to K^+\pi^-]$) the effects of such mixing 
will be at most $O(10\%)$ on the rates (i.e. $d(k,X)$). 

In particular there are two possible ways to deal with mixing

\begin{enumerate}

\item Using information on the the time between the $B^-$ decay and the
subsequent $D^0$ decay, then the effects of possible mixing can be
eliminated. 

\item
If the parameters of $D\xba D$ mixing are known independently, then they 
can be taken into account in interpreting the time integrated data

\end{enumerate}

\noindent
Indeed, if the mixing parameters and time dependent data is available, 
then one can in principle extract $\gamma$ from just one mode though most 
likely, the time dependence in the decay is too weak to make this a 
useful method.

In order to examine the question, let us first define the standard 
parameterization of this mixing as described 
in~\cite{ddbar_gen}. 
We denote the mass eigenstates as

\begin{eqnarray}
|D_L\rangle =p|D^0\rangle +q|\xba D^0\rangle
~~~~
|D_H\rangle =p|D^0\rangle -q|\xba D^0\rangle
\label{ddmixing1}
\end{eqnarray}

\noindent
where $D_L$ and $D_H$ represent the light and heavy eigenstates respectively.
If we denote the ratio

\begin{eqnarray}
q/p=r_{qp}e^{2i\phi}
\end{eqnarray}

\noindent
where $\phi$ is a CP violating phase which could, in principle, be 
present in $D^0-\xba D^0$ mixing and CP is also violated if 
$r_{qp}\neq 1$ though here we will suppose that $r_{qp}\approx 1$ as 
would likely be the case for CP violation generated by physics beyond the 
SM.

Let us denote by $m_H$ and $m_L$ the mass of the $D_H$ and $D_L$ states 
respectively. Likewise, we denote by $\Gamma_H$ and $\Gamma_L$ the widths 
of these states. In terms of these quantities, we will use the definitions:

\begin{eqnarray}
&&m={m_H+m_L\over 2},
~~~~~
\Gamma={\Gamma_H+\Gamma_L\over 2},
~~~~~
\mu=m-i\Gamma/2,
\nonumber\\
&&\Delta m = m_H-m_L,
~~~~~
\Delta\Gamma=\Gamma_H-\Gamma_L,
\nonumber\\
&&\xd={\Delta m\over\Gamma},
~~~~~
\yd={\Delta\Gamma\over 2\Gamma},
~~~~~
\zd=\xd+i\yd \equiv Z e^{i\lambda},
\label{ddmix_defs}
\end{eqnarray}

\noindent 
and denote by $|D_{phys}(t)>$ and $|\xba D_{phys}(t)>$  the 
time evolved state which is created at $t=0$ as $D_{phys}^0$ and
$\xba D_{phys}^0$ respectively. These are thus related to the flavor 
eigenstates as:

\begin{eqnarray}
|D_{phys}(t)>      &=&
g_+(t)|D^0> + r_{qp}e^{2i\phi}|\xba D^0>;~~
\nonumber\\
|\xba D_{phys}(t)> &=&
g_+(t)|\xba D^0> + r_{qp}^{-1}e^{-2i\phi}|D^0>
\end{eqnarray}

\noindent
with $g_\pm$ are given by:

\begin{eqnarray}
g_+(t)=e^{-i\mu t}\cosh({i\over 2}  \zd \tau);
~~~~~
g_-(t)=e^{-i\mu t}\sinh( {i\over 2} \zd \tau).
\end{eqnarray}

\noindent
with $\tau=\Gamma_D t$.

A number of experiments have recently produced results which bound these
mixing parameters.  The E791 experiment at Fermilab
obtains\cite{zdlimit_e791} $\Delta\Gamma=0.04\pm0.14~ps^{-1}$
corresponding to $y=0.8\pm 2.9\pm 1.0\%$ from the study of the life time
ratio of $D^0\to K^-\pi^+$ versus $D^0\to K^+K^-$. CLEO has
reported\cite{zdlimit_cleo} a 95\% c.l. bound of $|x^\prime|\leq 2.8\%$
and $-5.8\%<y^\prime<1.0\%$ where $x^\prime = x\cos\delta_{K^+\pi^-} +
y\sin\delta_{K^+\pi^-}$ and $y^\prime = -y\sin\delta_{K^+\pi^-} +
x\cos\delta_{K^+\pi^-}$ obtained through the study of the time dependence
of the decay $D^0\to K^+\pi^-$. Results from the FOCUS
experiment\cite{zdlimit_focus} at Fermilab, again using the lifetime
ratio, give $y=3.42\pm 1.39\pm 0.74\%$ which interestingly is 2.5-$\sigma$
from 0.

In the Standard Model, $D^0-\xba D^0$ mixing receives contributions from
both short distance and long distance processes.  The short distance
mixing via the box diagram may be reliably estimated to be
$\yd<\xd=O(10^{-5})$. On the other hand, the long distance effects involve
considerable uncertainty from the hadronic interactions involved. 
Calculations based on dispersion theory~\cite{ddmix_disp} suggest that
$|\zd|<10^{-4}$. On the other hand it has recently been
suggested~\cite{ddmix_resonance} that life time differences driven by
kaonic resonances in the vicinity of the $D^0$ mass could lead to $\yd\sim
10^{-2}$.  In the Standard Model, the CP violating phase $\phi$ will in
all cases be negligible.

It is also possible that $D^0-\xba D^0$ mixing is driven by new physics
beyond the standard model. Many examples of such mixing have been
considered (see compilation in \cite{ddmix_theory}) and in this case,
additional CP violation through the phase angle $\phi$ is also possible.

In the case we are interested in, $B^-\to K^- [D^0\to X]$, the initial
decay is to a mixed state of $D^0-\xba D^0$ where both the $D^0$ and $\xba
D^0$ components have some amplitude to decay to $X$. Since X is small, it
is justified to expand the mixing effects on the time dependent decay as
follows: 

\begin{eqnarray}
{d\over d\tau}d(k,X)&\approx&(d_0(k,X)+d_1(k,X)\tau)e^{-\tau}
\nonumber\\
{d\over d\tau}\xba d(k,\xba X)
&\approx& 
(\xba d_0(k,\xba X)
+
\xba d_1(k,\xba X)\tau)e^{-\tau}
\end{eqnarray}

\noindent Since $d_0$ and $\xba d_0$ represent the decay rate at 0 time
interval, they will be identical with the expressions given in
eq.~(\ref{firstmode}). If we assume that there is no CP violation in the
$D\xba D$ system, then $d_1$ and $\xba d_1$ are given by:

\begin{eqnarray}
{1\over 2}(d_1+\xba d_1)
&=&
+2\sqrt{ac(\xba X)}Z
\left\{
\sqrt{a c(X)}\sin(\lambda+\delta)\cos\gamma
+
\sqrt{b c(\xba X)}\sin(\lambda-\zeta)\cos(\gamma+\phi)
\right \}
\nonumber\\
&&+
2\sqrt{bc(X)}Z
\left\{
\sqrt{b c(\xba X)}\sin(\lambda-\delta)\cos\gamma
+
\sqrt{a c(X)}\sin(\lambda+\zeta)\cos(\gamma+\phi)
\right \}
\nonumber\\
\nonumber\\
{1\over 2}(d_1-\xba d_1)
&=&
-2\sqrt{ac(\xba X)}Z
\left\{
\sqrt{ac(X)}\cos(\lambda+\delta)\sin\gamma
+
\sqrt{bc(\xba X)}\cos(\lambda-\zeta)\sin(\gamma+\phi)
\right\}
\nonumber\\
&&+
2\sqrt{bc(X)}Z
\left\{
\sqrt{bc(\xba X)}\cos(\lambda-\delta)\sin\gamma
+
\sqrt{ac(X)}\cos(\lambda+\zeta)\sin(\gamma+\phi)
\right\}
\nonumber\\
\label{timedep}
\end{eqnarray}

In order to estimate the relative effect of the mixing, let us assume 
that $Z\approx 0.01$ 
and recall from our rough estimates that 

\begin{eqnarray}
{b\over a} 
\approx 
{c(X) \over c(\xba X)}
\approx
0.01
\end{eqnarray}

\noindent
and thus from eq.~(\ref{timedep}) 
we can estimate:

\begin{eqnarray}
{d_1\over d_0}
\approx
{\xba d_1\over \xba d_0}
\approx
0.1
\label{tdep_est}
\end{eqnarray}

Note that in Eqn.~(\ref{timedep}) the time dependent term depends on both
$\zeta$ and $\delta$ rather than just the sum $\xi$ and, in addition, it
depends on the mixing parameters.  The simplest way to extract $\gamma$
given time dependent data is therefore to try and obtain the values of
$d_0$ and $\xba d_0$ which do not have these extra dependences reducing
the problem to the same situation as if mixing were not present. 

This can be accomplished through weighting the data with 

\begin{eqnarray}
w_0(\tau)
=
2-\tau
\end{eqnarray}

so that

\begin{eqnarray}
d_0=
\int_0^\infty	 
[d(\tau)w_0(\tau)] d\tau 
;
~~~
\xba d_0=
\int_0^\infty
[\xba d(\tau)w_0(\tau)] d\tau
\label{timeweight}
\end{eqnarray}

Using this method more data would be required to obtain the same
statistical results as in the unmixed case. In the unmixed case where
$d_1=0$ one could obtain $d_0$ more effectively by taking the time
integrated rate. Thus in the unmixed case if a measurement of $d_0$ is
based on $n$ events, the uncertainty in $d_0$ is given by:

\begin{eqnarray}
{(\Delta d_0)^2\over d_0^2}
=
{1\over n}~~~~~(no~mixing)
\end{eqnarray}

\noindent In the mixed case, using eqn.~(\ref{timeweight}) the uncertainty
is

\begin{eqnarray}
{(\Delta d_0)^2\over d_0^2}
=
{2\over n}\left(1+{d_1\over d_0}\right)^2 ~~~~~(with~mixing)
\end{eqnarray}

\noindent From Eqn.~(\ref{tdep_est}) this means that roughly twice the
data is needed to have the same statistical power as in the unmixed case. 
In order to gauge the precision  
of time measurement required, we can smear out the distribution in 
eq.~(\ref{timeweight}) with a Gaussian resolution function of the form

\begin{eqnarray}
r(\tau,\tau^\prime)\propto e^{-{(\tau-\tau^\prime)^2\over 2\sigma^2}}
\end{eqnarray}

\noindent where $\tau$ is the actual time of the decay, $\tau^\prime$ is
the measured time of the decay and $\sigma$ is the resolution (all in
units of $1/\Gamma_D$).  Since $r$ is symmetric under $\tau\leftrightarrow
\tau^\prime$, the fact that $w$ is linear in $\tau$ implies
eq.~(\ref{timeweight}) will still be true for $\tau^\prime$ but now the
error is: 

\begin{eqnarray}
{(\Delta d_0)^2\over d_0^2}
=
{2\over n}(1+\sigma^2)\left(1+{d_1\over d_0}\right)^2 
\end{eqnarray}

\noindent
As can be seen, the number of events required is not adversely effected if 
$\sigma \leq 1/\Gamma_D$ but will be significantly degraded otherwise.

In the above strategy, it was assumed that the mixing was relatively
unconstrained except for the assumption that $d_1$ is smaller than $d_0$
which seems justified. If the rate of $D\bar D$ mixing is near the current
experimental bounds, however, then there is another possible way to
obtain $\gamma$.  If one measures $\{d_0$, $d_1$, $\xba d_0$, $\xba d_1\}$
and knows all of the mixing parameters, then there are just four unknown
parameters that need to be fitted for: $\{\gamma$, $\delta$, $\zeta$, and
$b\}$. We therefore have four equations in four unknowns and can solve for
$\gamma$.  To obtain $d_1$ we can use the weight function: 

\begin{eqnarray}
w_1(\tau)=\tau-1
\end{eqnarray}

It is also possible to learn about $\gamma$ 
using time integrated data.
This has been previously considered 
in~\cite{soffer_silva}. The corrections due to mixing in the 
approximation we are using are:

\begin{eqnarray}
d&=&d_0+d_1\nonumber\\
\xba d &=& \xba d_0 +\xba d_1
\end{eqnarray}

In the case that the $D\xba D$ mixing parameters are already determined,
if $n$ modes are measured, there are $3+n$ unknowns:  $\zeta$, $\gamma$,
$b$ and for each mode $X_i$, $\delta_i$. Each mode provides two pieces of
information, $d$ and $\xba d$, leading to a total of $2n$ measurements. To
solve for the unknowns at least 3 modes are thus required (six equations
in six unknowns). If the mixing parameters are unconstrained, this adds
four more unknowns: $\{x,y,r_{qp},\phi\}$ and in principle at least seven
modes would be required.

On the other hand, using such time dependent information is probably not
the best way to measure $D\xba D$ oscillation parameters.  Perhaps the
best way to proceed is to consider $d_1$ and $\xba d_1$ to be
corrections to the time integrated data. From eqn.~(\ref{timedep}) one can
take the bound on $x$ and $y$ together with an estimate of $b$ and bound
these terms. As indicated, this gives about a $10\%$ correction which, as
discussed in \cite{soffer_silva} leads to about a $15^\circ$ error in
$\gamma$.  This is similar to the errors typically obtained in the unmixed
case for $\hat N_B=10^8$ and so to gain improvements in the determination
$\gamma$ with larger numbers of $N_B$ in this scenario one would both have
to improve the bounds on $D\xba D$ mixing (or indeed discover it). Of
course since measurements of $D\xba D$ mixing can often be made at a $B$
facility, it would be natural for such improvements to occur at the same
time.

\newpage

\begin{table}[htb]
\begin{center}
\caption{Cabibbo allowed (CBA) and doubly Cabibbo suppressed (DCS)
modes of $D^0 (\xba D^0)$. $BR$'s for CBA modes are taken from 
PDB \protect\cite{pdb}
while
those for DCS are predicted, using the model given in the text, except
for $\xba D^0 \to K^+\pi^-$ wherein the measured value is shown.
\label{tabone}}
\bigskip
\bigskip
\begin{tabular}{|l|c|c|}
\hline
Mode & $Br(D^0 \to {\rm final\, state})$ & $Br (\xba D^0\to {\rm
final\, state})$ \\ 
\hline
\hline
$K^+\pi^-$ & $(2.9\pm1.4)\times 10^{-4}$ & $3.83\times10^{-2}$ \\
\hline
$K^+\rho^-$ & $3.8\times 10^{-4}$ & $ 10.8\times10^{-2}$ \\
\hline
$K^+a^-_1$ & $7.0\times 10^{-5}$ & $7.3\times 10^{-2}$ \\
\hline
$K^{\ast+}\pi^-$ & $8.3\times10^{-4}$ & $5.0\times 10^{-2}$\\ 
\hline
\end{tabular}
\end{center}
\end{table}

\bigskip

\begin{table}
\begin{center}
\caption{
The branching ratios for the combined decay $B^-\to K^{*-} D^0$ followed 
by the decay of $D^0$ to the modes given below using the parameters 
considered in Section~6 with $\gamma=60^\circ$ and the given strong 
phases $\xi_i$. $d_i$ and $\xba d$ are given in units of $10^{-8}$ and 
$\alpha^\prime$ is the 
partial rate asymmetry.
\label{tabtwo}}
\bigskip
\bigskip
\begin{tabular}{|c|c|c|c|c|c|}  
  \hline
&&&&&\\
  Mode & $d_i$ & $\overline {d_i}$ & ${1\over 2}(d_i+\overline{d_i})$ &
  $\alpha^\prime$  & $\xi_i$  \\
  \hline
  $K^+ \pi^-$  &      91 &      75 &      83  &   
0.096 &      10 \\
  \hline
  $K_s \pi^0$  &     842 &     740 &     791  &   
0.064 &      20 \\
  \hline
  $K^+ \rho^-$ &     289 &     159 &     224  &   
0.288 &      30 \\
  \hline
  $K^+ a_1^-$  &     203 &      90 &     146  &   
0.383 &      40 \\
  \hline
  $K_s \rho^0$ &     333 &     391 &     362  &   
0.081 &     200 \\
  \hline
  $K^{*+}\pi^-$ &      97 &      34 &      65  &   
0.477 &      50 \\
  \hline
\end{tabular}
\end{center}
\end{table}

\bigskip

\begin{table}
\begin{center}
\caption{
The parameters of the model for $D^0\to K^-\pi^+\pi^0$ decay obtained 
in~\cite{e687ref}. 
\label{tabthree}}
\bigskip
\bigskip
\begin{tabular}{|c|c|c|}
\hline
channel & decay fraction (\%) & $\delta_r-\delta_\rho$ (deg.) \\
\hline
non-resonant & $10.1\pm 3.3\pm 3.0\pm 2.7$  
&  $-122\pm 10\pm 21\pm 2$ \\
$K^{*0}$      & $16.5\pm 3.1\pm 1.1\pm 1.1$  
&  $-2  \pm 12\pm 23\pm 2$ \\
$K^{*+}$    & $14.8\pm 2.8\pm 4.9\pm 0.3$  
&  $162 \pm 10\pm 7 \pm 4$ \\
$\rho^-$    & $76.5\pm 4.1\pm 2.2\pm 4.9$  
&  $0$    \\
\hline
\end{tabular}
\end{center}
\end{table}

\bigskip

\newpage
\flushleft{\bf\large Figure Captions}

\bigskip
\bigskip
{\bf Figure 1:}
Each of the solid lines shows the locus of points in $\gamma$
versus $u_i$ of allowed solutions given $z_i=1.5$ for $y_i=0$ (outer 
curve), $1$ (intermediate curve) and $2$ (inner curve). The boxes 
indicate the inequalities Eqns.~(\ref{gammalimit},\ref{ulimit}).

\bigskip {\bf Figure 2:} The solid line shows the allowed points on the
$\gamma-b(K^*)$ plot given a measurement of $B^-\to K^* D^0$ followed by
$D^0\to K^+\pi^-$ (and charge conjugate) using the estimated values of
Section 5 eqns.~(\ref{aest},\ref{best}) assuming $\gamma=90^\circ$. The
dashed line shows the results which would follow in the same case given a
measurement of $D^0\to K_s\pi^0$ in the final state.

\bigskip
{\bf Figure 3:}
(a)~The likelihood distribution is shown 
as a function of $\gamma$ and 
$b(K^*)$ 
assuming that $\nup=10^8$ with the 
branching ratios considered in Table~2 and assuming only the $K^+\pi^-$ 
and $K_s\pi^0$ modes are measured. 
The outer edge of the 
the shaded regions correspond to 
$90\%$  confidence 
while the inner edge corresponds to 
$68\%$  confidence. The solid lines 
show 
the locus of points which give exactly the $K^+\pi^-$ results while the 
short dashed curve shows the points which give the 
$K_s\pi^0$ results. 

(b)~The likelihood distribution as in Fig.~3a is shown assuming all of the 
modes in Table~2 are used.
The solution for the $K^+\pi^-$ data is shown with the solid curve;
that for the $K_S\pi^0$ data is shown with the short dashed curve;
the one for the $K^+\rho^-$ data is shown with the long dashed curve;
the one for the $K^+a_1^-$ data is shown with the dash-dot curve;
the one for the $K_s \rho^0$ data is shown with the dash-dot-dot 
curve  and
the solution for the $K^{*+} \pi^-$ data is shown with the dash-dash-dot 
curve.

\bigskip
{\bf Figure 4:}
The ratio between the 
the likelihood distribution 
and the maximum likelihood
is shown as a function of $\gamma$ with the 
parameters as in Fig.~3b except $\gamma$ is taken to be 
$15^\circ$ (dashed curve);
$30^\circ$ (solid curve);
$60^\circ$ (dotted curve);
$90^\circ$ (dash-dot curve).

\bigskip
{\bf Figure 5:}
(a)~Dalitz plots for the decay products of the $D^0$ in the decay chain 
$B^-\to K^{*-} D^0$ followed by $D^0\to K^+\pi^-\pi^0$ (upper plot) and 
the charge conjugate process (lower plot). We use the model described in 
section~\ref{section_6} 
where we take $\gamma=90^\circ$ and $\xi=90^\circ$. The plots 
represent the results given $N_B=10^{10}$.

(b)~Dalitz plot for the same system as in Fig.~5(a) with $\xi=0^\circ$.

\bigskip
{\bf Figure 6:}
(a)~A plot of the partial rate asymmetry for the system in Fig.~5 as a 
function of $\gamma$.

(b)~The $N_B$ required for a $3-\sigma$ signal as a function of $\xi$ 
using only the PRA (solid line) and using the optimal observable (dashed 
line).

\bigskip 
{\bf Figure 7:} 
(a)~The locus of points on a dalitz plot where $Q_{min}=Q$ for
$\gamma=60^\circ$ and $\zeta=0^\circ$ (solid line); $\zeta=30^\circ$
(dashed line); $\zeta=60^\circ$ (dotted line) and $\zeta=90^\circ$
(dot-dashed line). 

(b)~A plot of $f(rQ)$ as a function of $\gamma$  
for $\zeta=0^\circ$ (solid curves); 
for $\zeta=30^\circ$ (dashed curves); 
for $\zeta=60^\circ$ (dotted curves) and 
for $\zeta=90^\circ$ (dash-dot curves). 
In each case the lower curve corresponds to 
$r=0.9$ 
and the upper curve to 
$r=0.7$.

(c)~A plot of $f^2(q)$ is shown as a function of $q$ for $Q=3/4$ (solid
line); $Q=1/2$ (dashed line) and $Q=1/4$ (dotted line) where
$\zeta=90^\circ$.

\bigskip
{\bf Figure 8:} A plot of $g^2_{min}(b)$ (solid line) and $g^2_{max}(b)$
(dashed line) line is shown as a function of $b/b(k)$. Here we have taken
$Q=3/4$ and $\zeta=\pi/2$.

\newpage
\begin{figure}
\epsfxsize 5.0 in
\mbox{\epsfbox{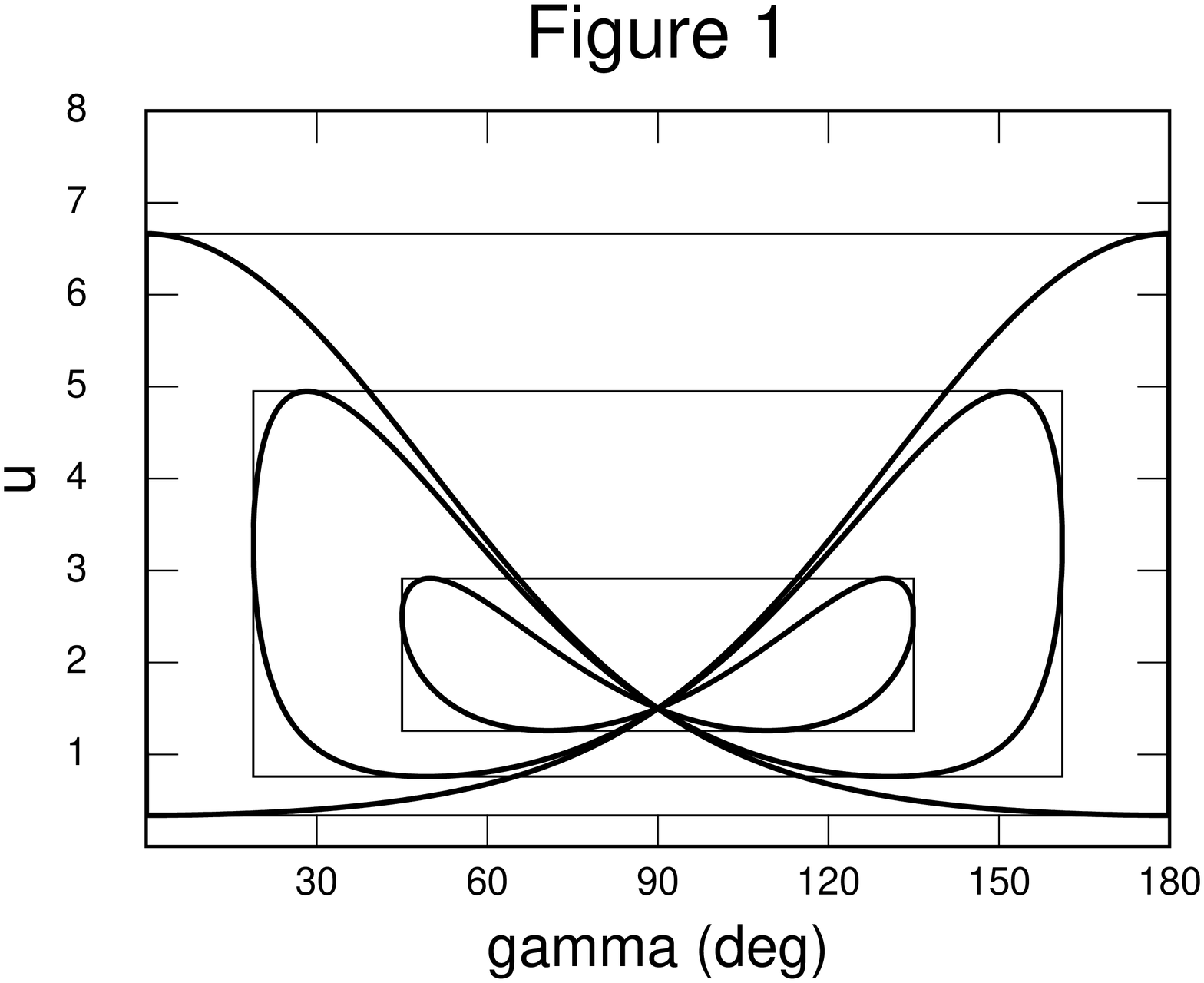}}
\end{figure}

\newpage
\begin{figure}
\epsfxsize 5.0 in
\mbox{\epsfbox{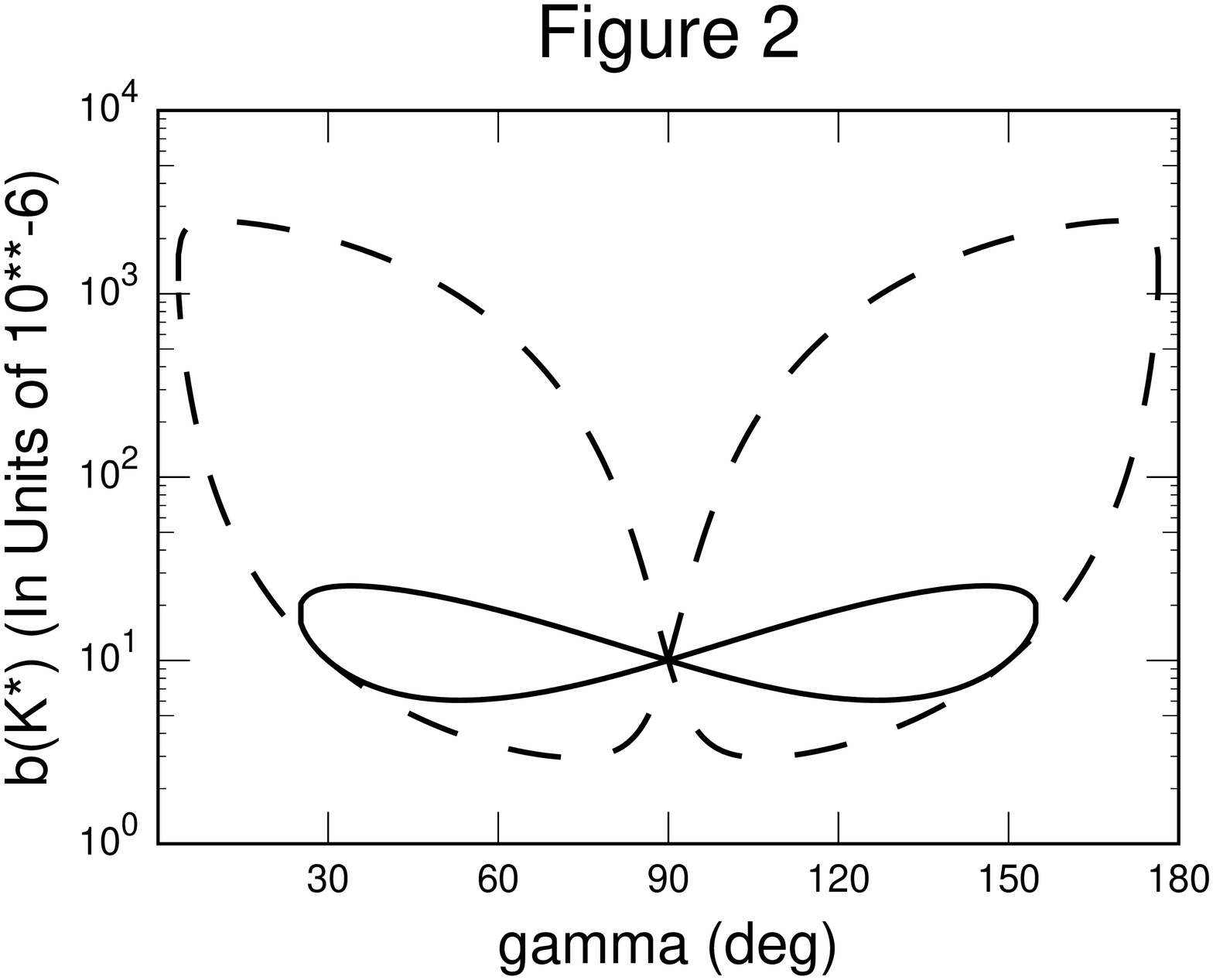}}
\end{figure}

\newpage
\begin{figure}
\epsfxsize 5.0 in
\mbox{\epsfbox{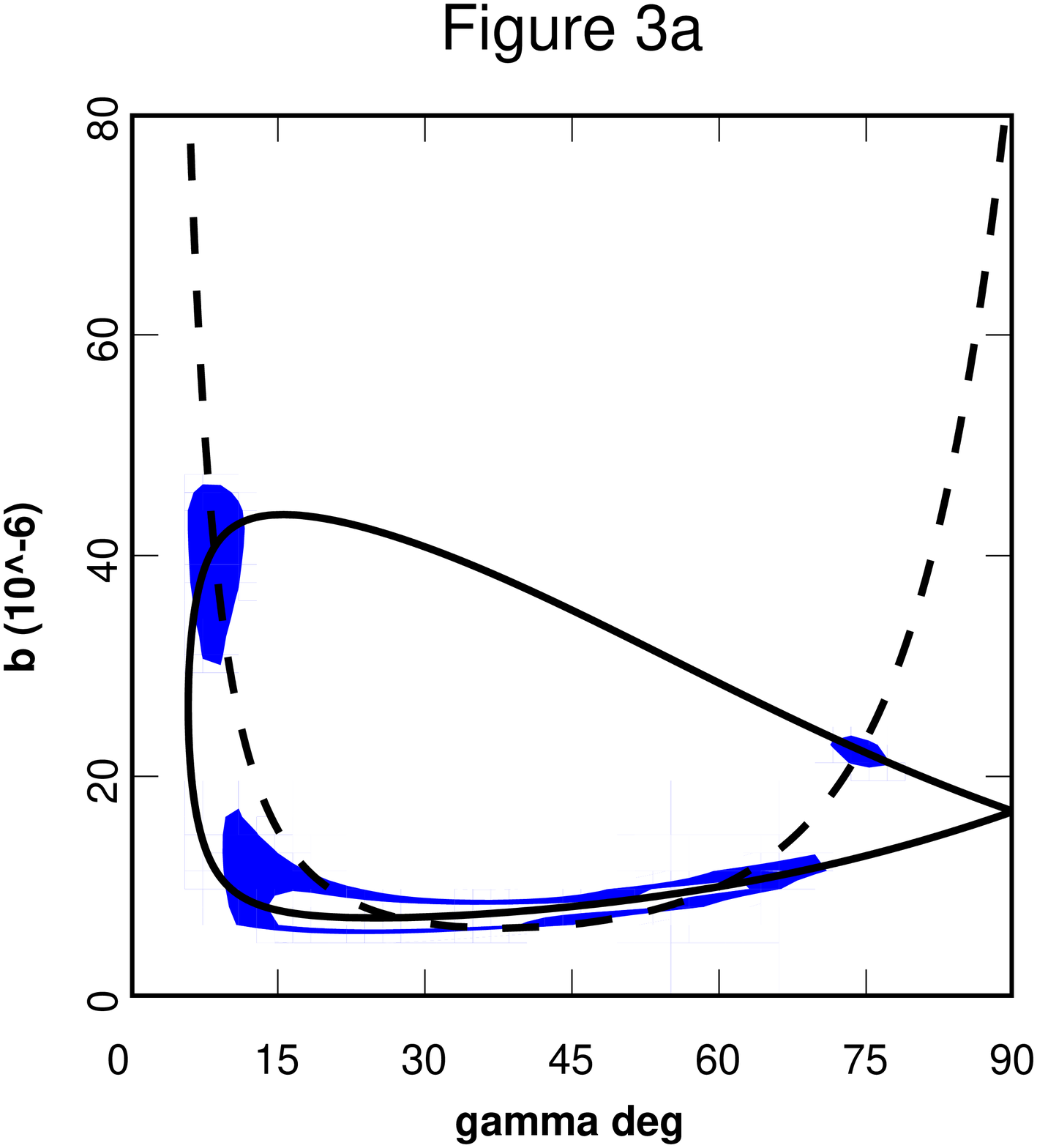}}
\end{figure}

\newpage
\begin{figure}
\epsfxsize 5.0 in
\mbox{\epsfbox{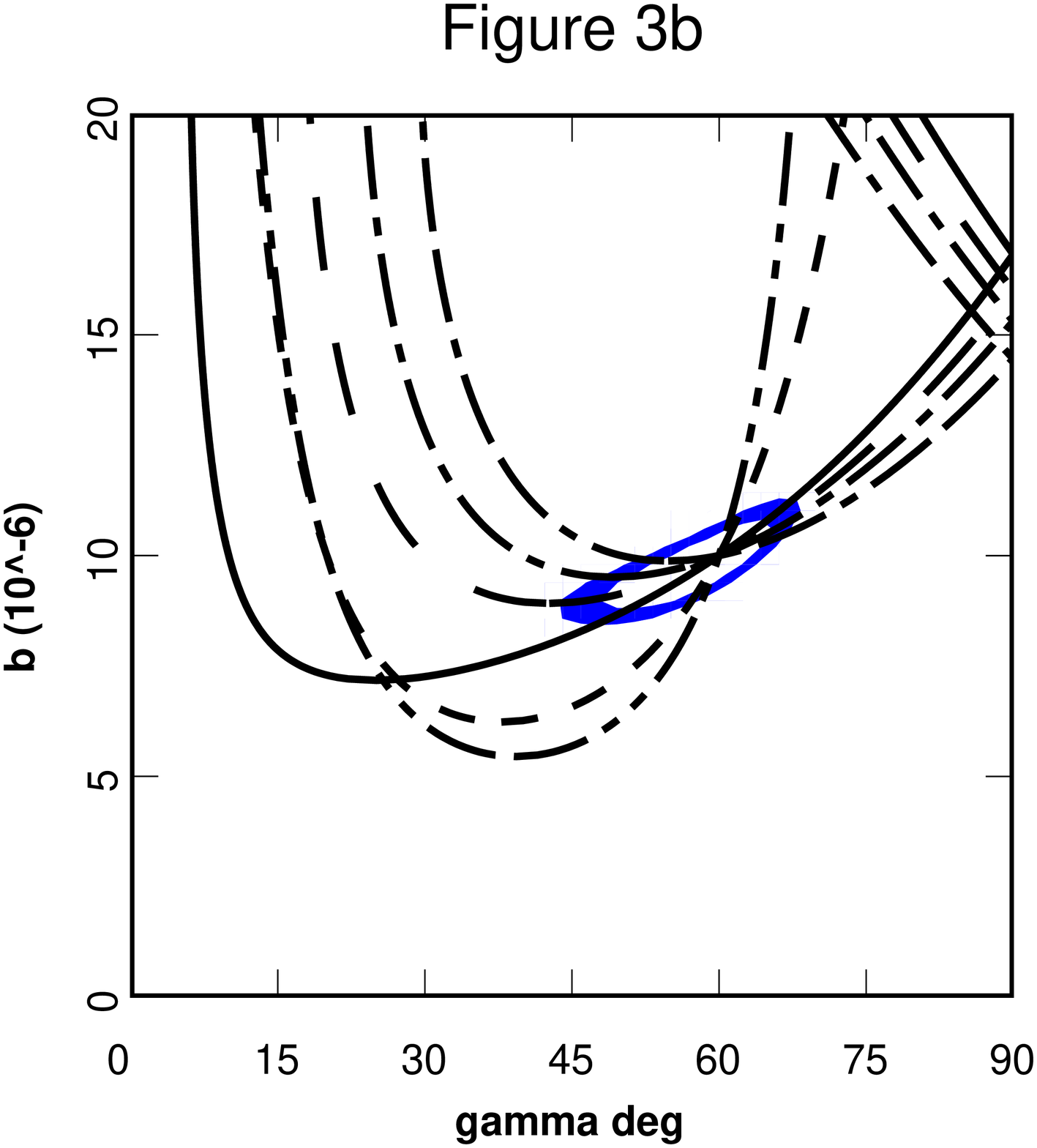}}
\end{figure}

\newpage
\begin{figure}
\epsfxsize 5.0 in
\mbox{\epsfbox{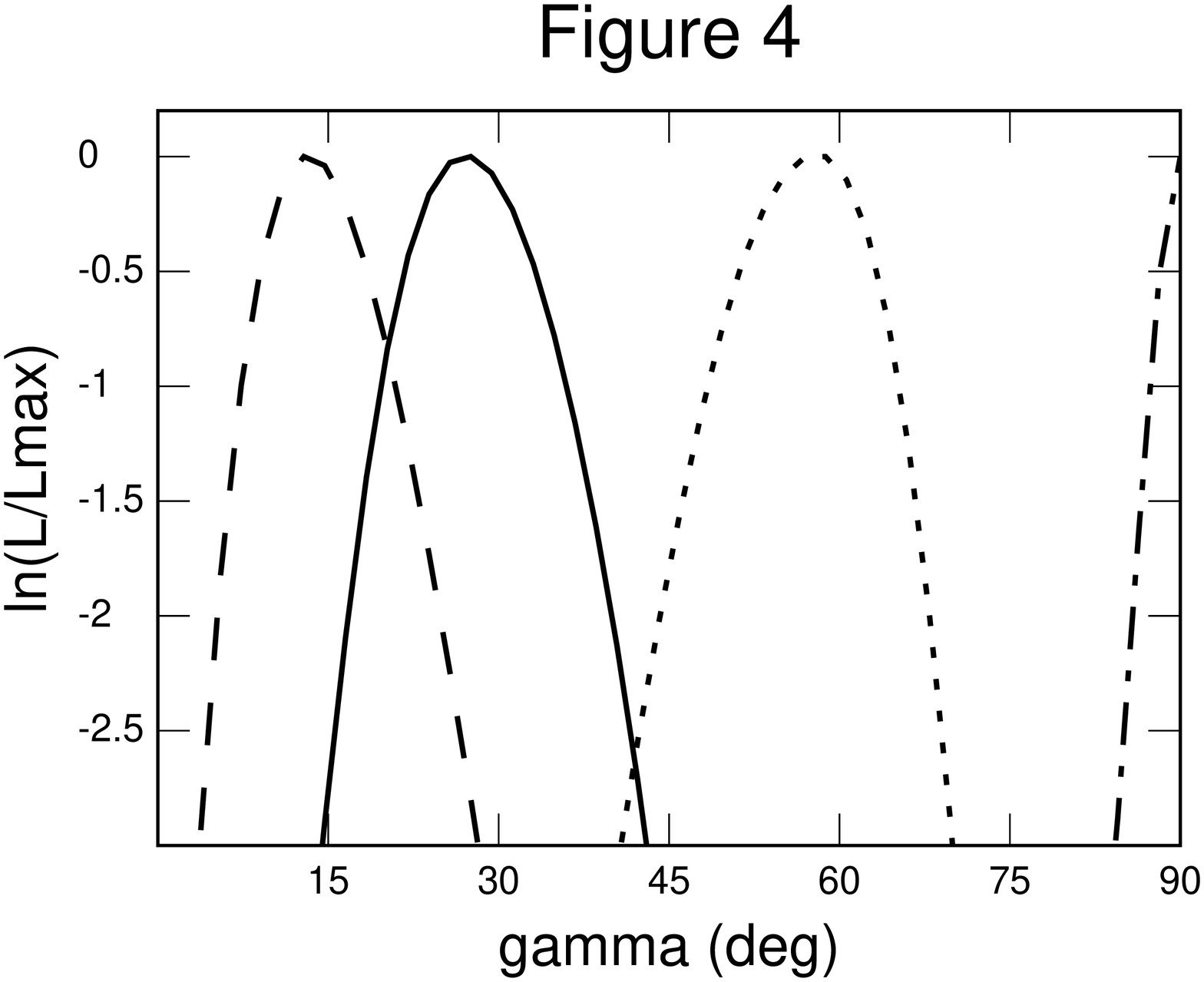}}
\end{figure}

\newpage
\begin{figure}
\epsfxsize 4.0 in
\mbox{\epsfbox{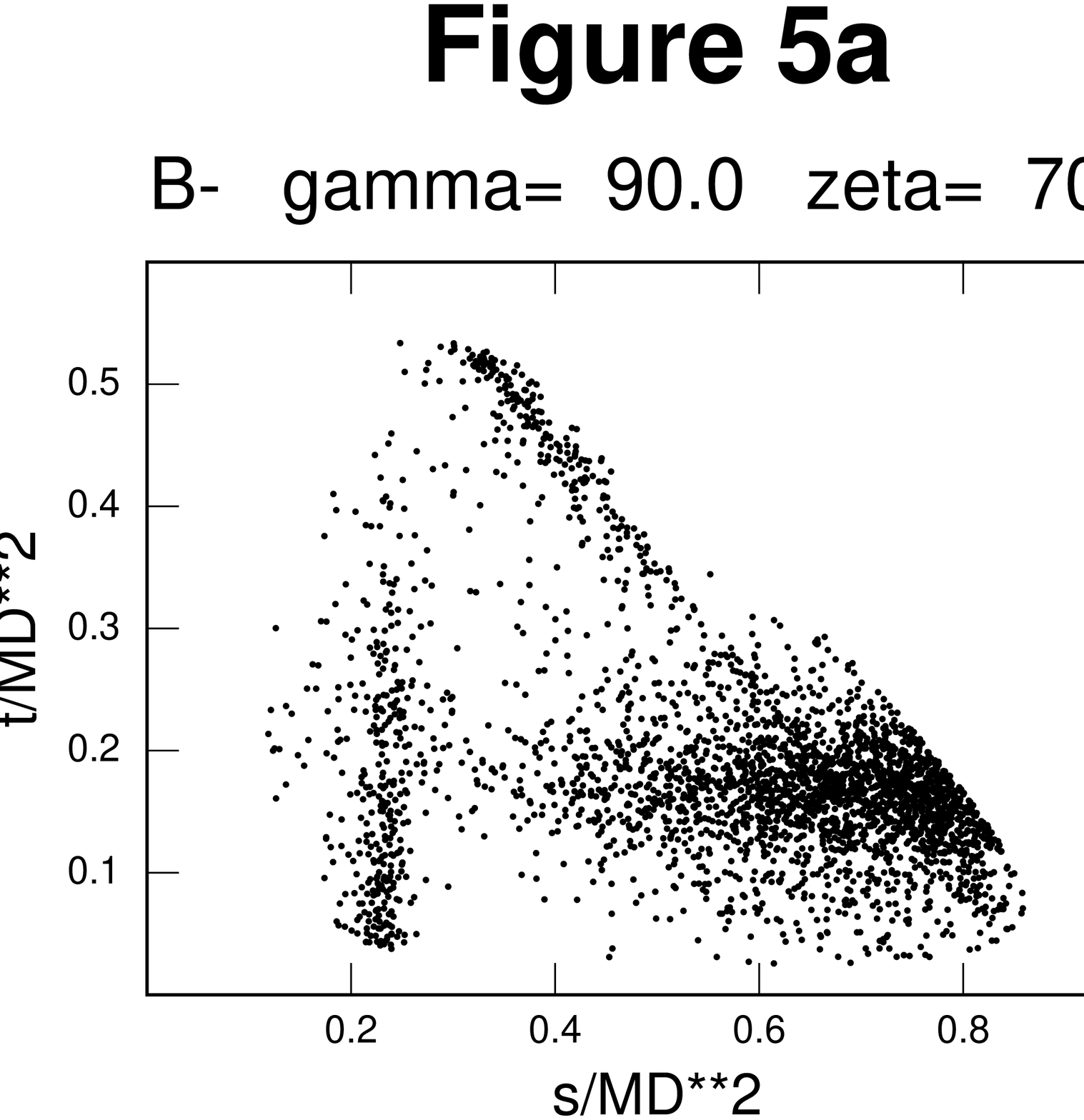}}
\end{figure}
\begin{figure}
\epsfxsize 4.0 in
\mbox{\epsfbox{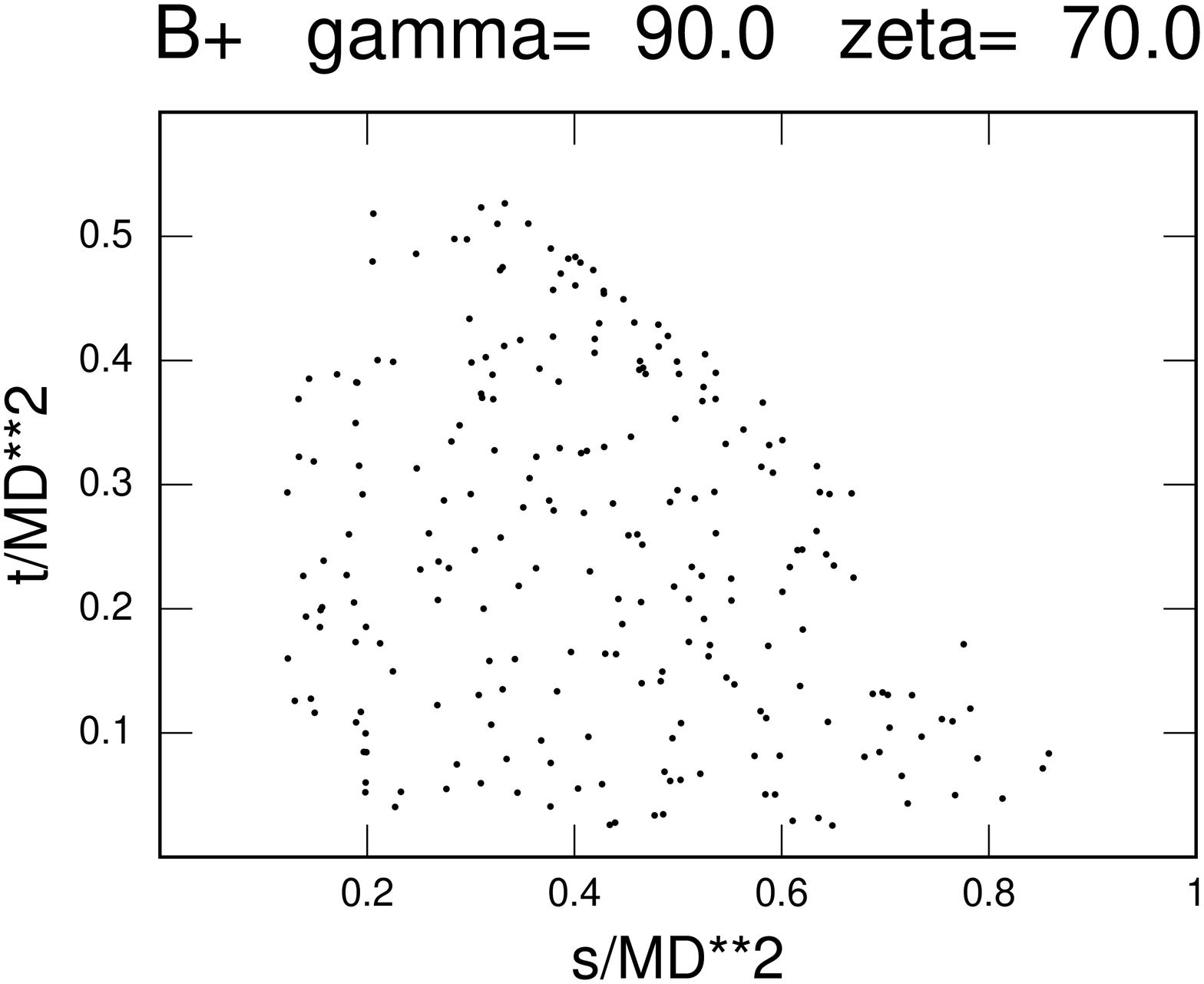}}
\end{figure}

\newpage
\begin{figure}
\epsfxsize 4.0 in
\mbox{\epsfbox{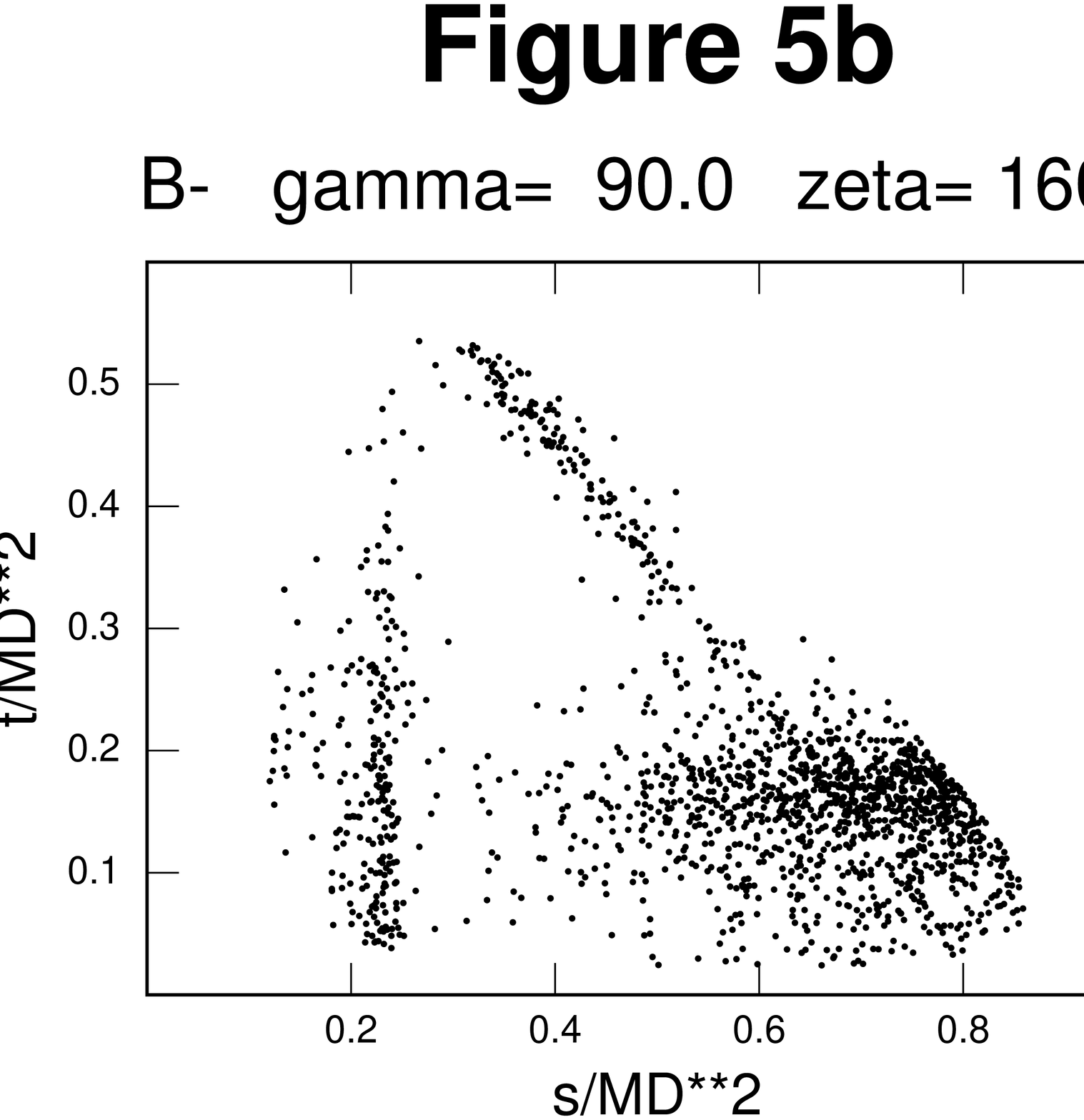}}
\end{figure}
\begin{figure}
\epsfxsize 4.0 in
\mbox{\epsfbox{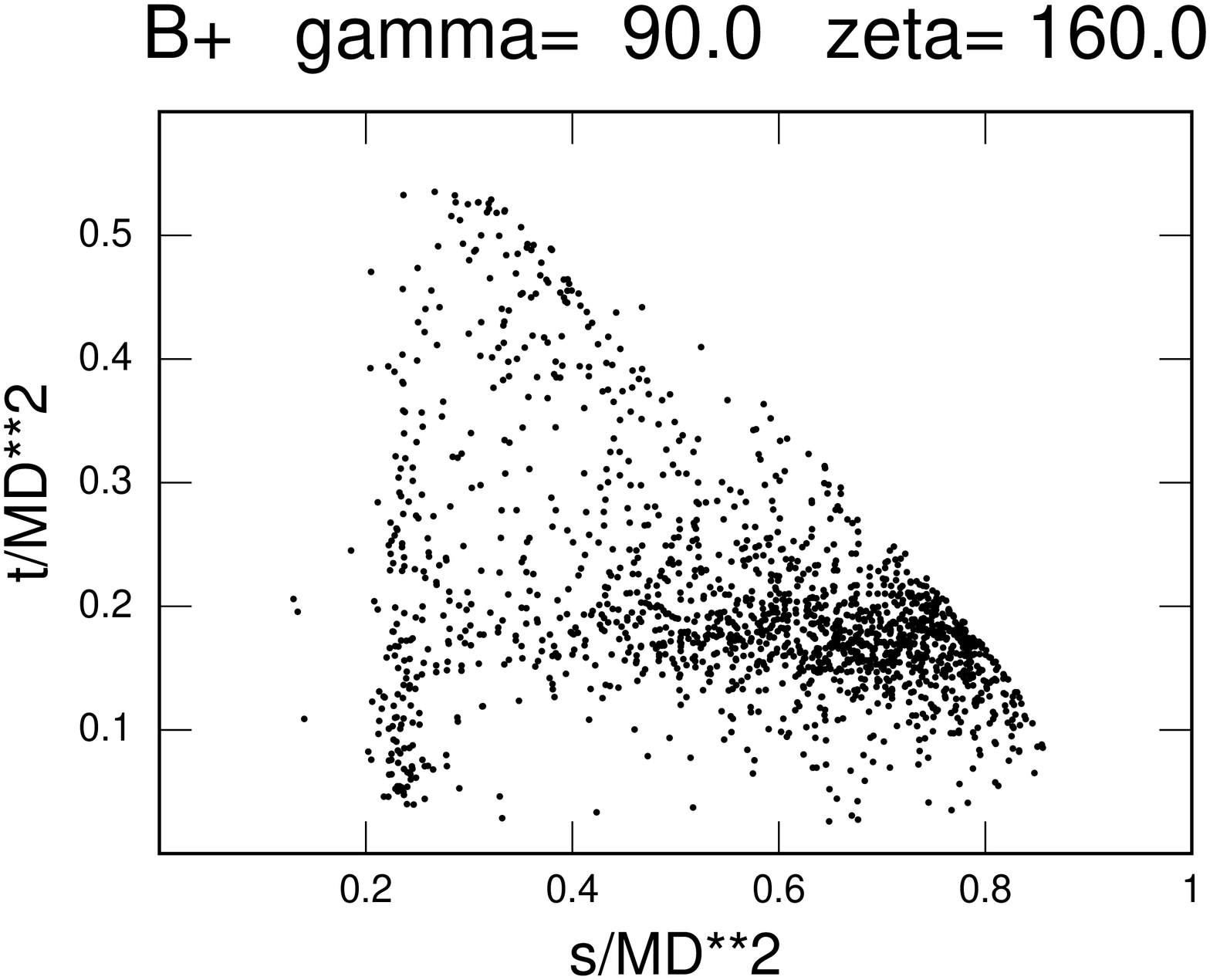}}
\end{figure}

\newpage
\begin{figure}
\epsfxsize 4.0 in
\mbox{\epsfbox{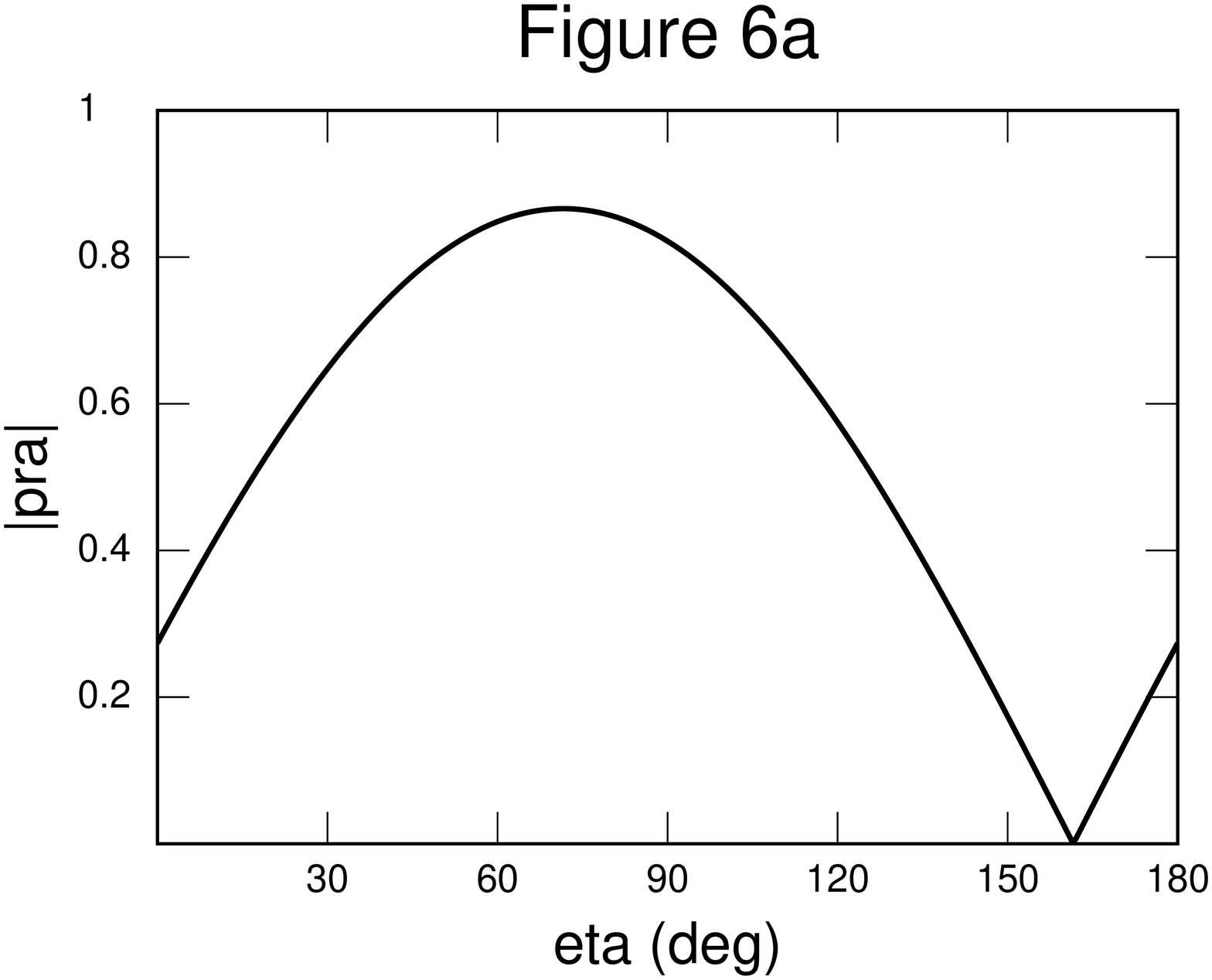}}
\end{figure}
\bigskip
\begin{figure}
\epsfxsize 4.0 in
\mbox{\epsfbox{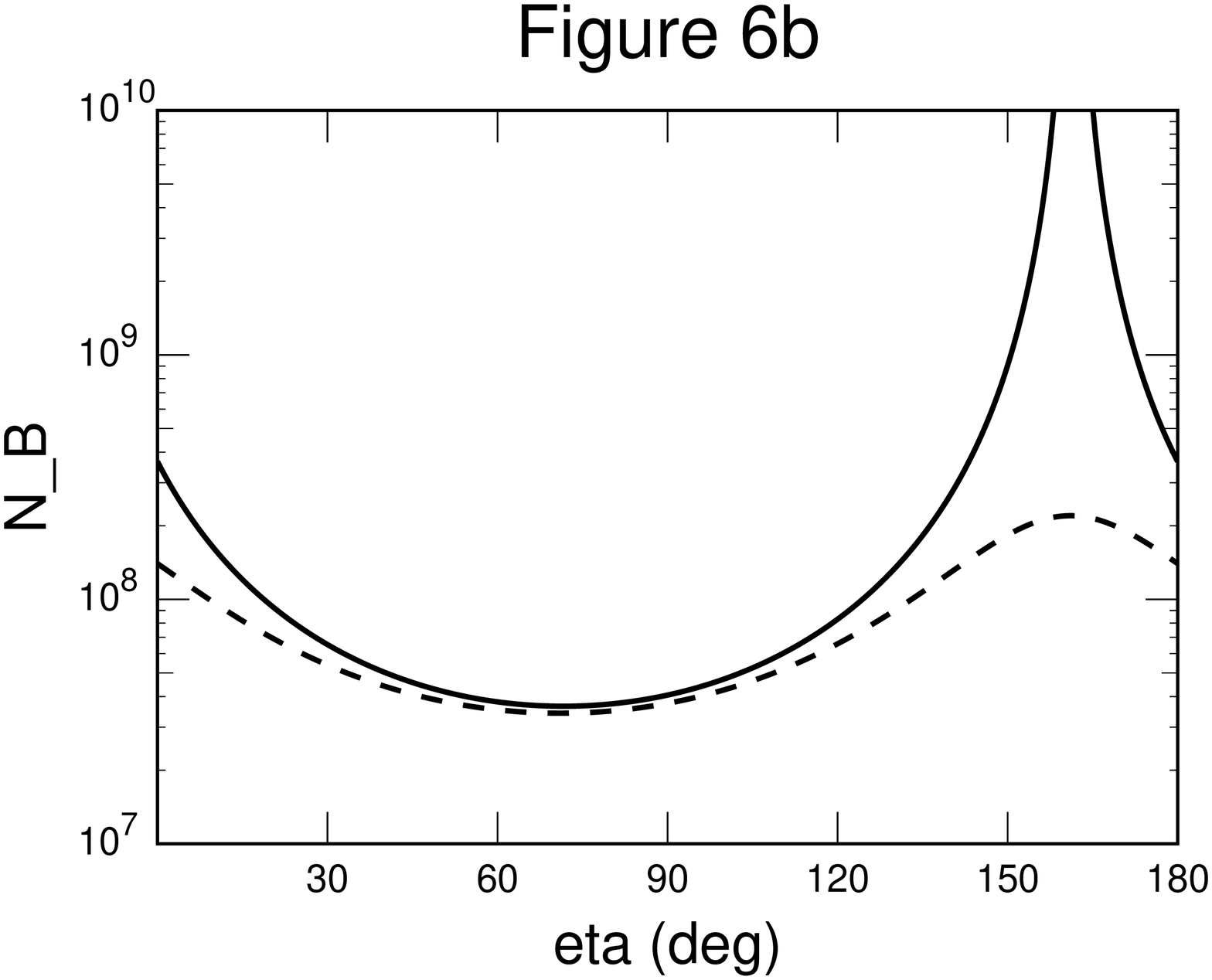}}
\end{figure}

\newpage
\begin{figure}
\epsfxsize 4.0 in
\mbox{\epsfbox{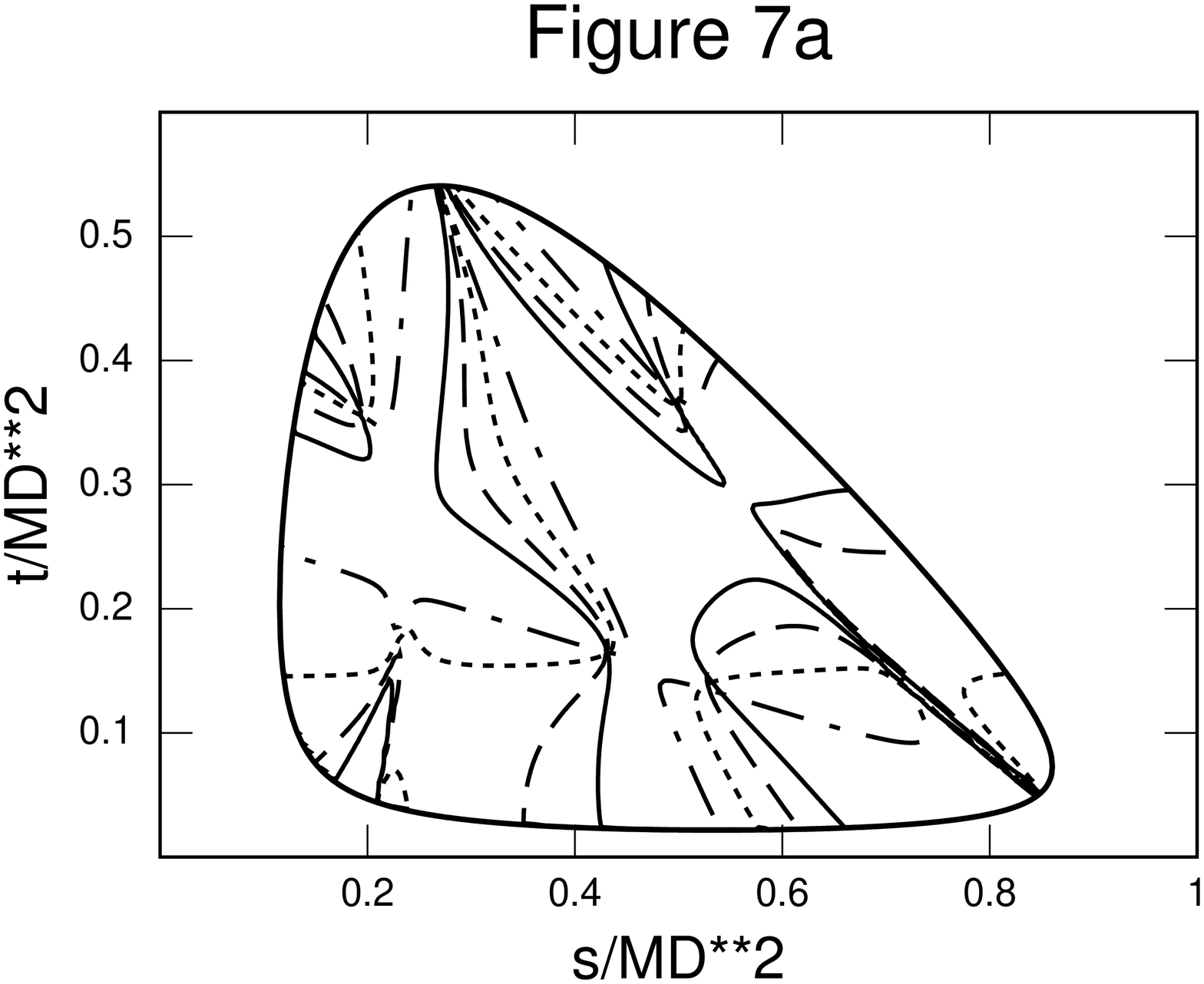}}
\end{figure}

\newpage
\begin{figure}
\epsfxsize 4.0 in
\mbox{\epsfbox{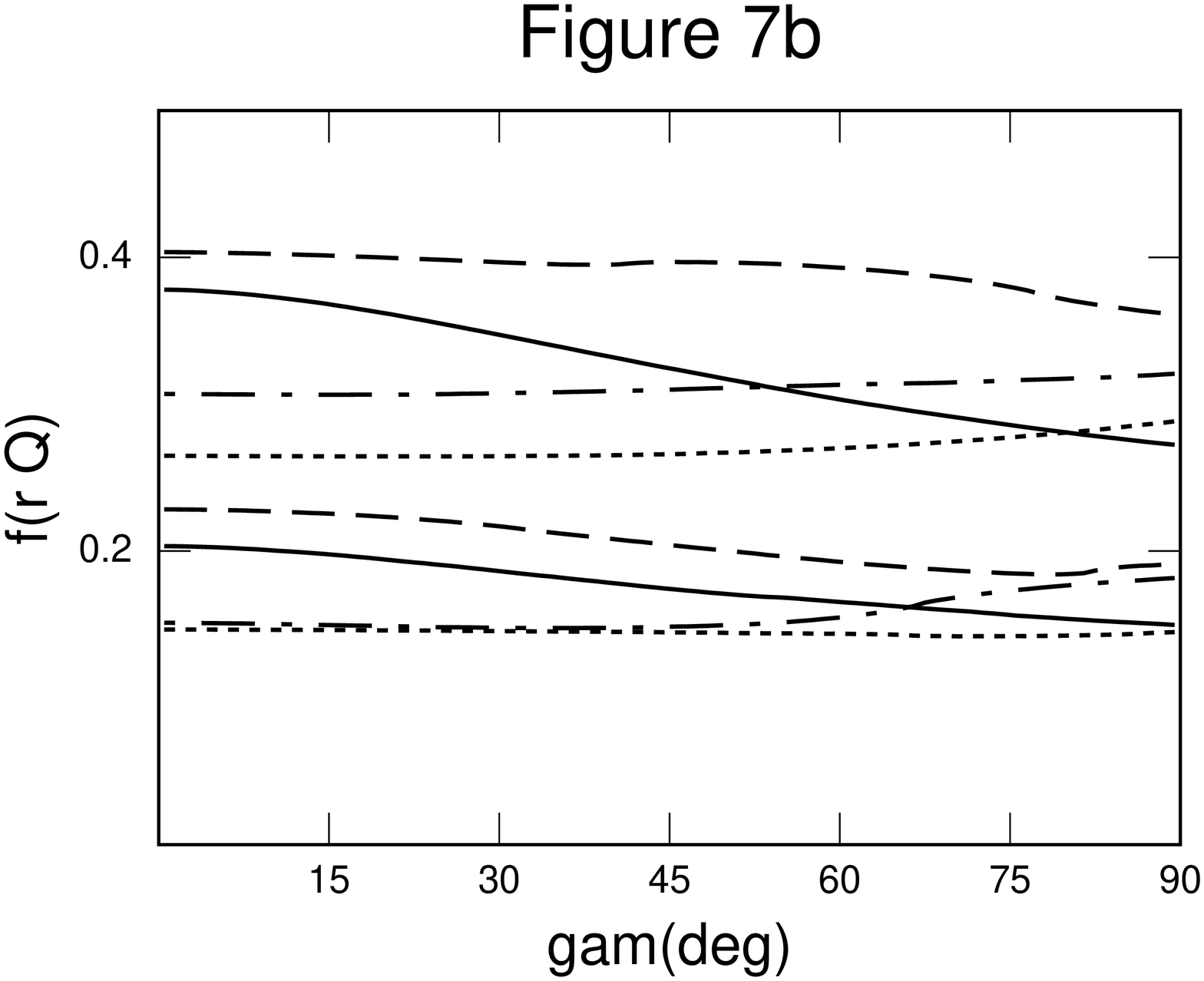}}
\end{figure}

\newpage
\begin{figure}
\epsfxsize 4.0 in
\mbox{\epsfbox{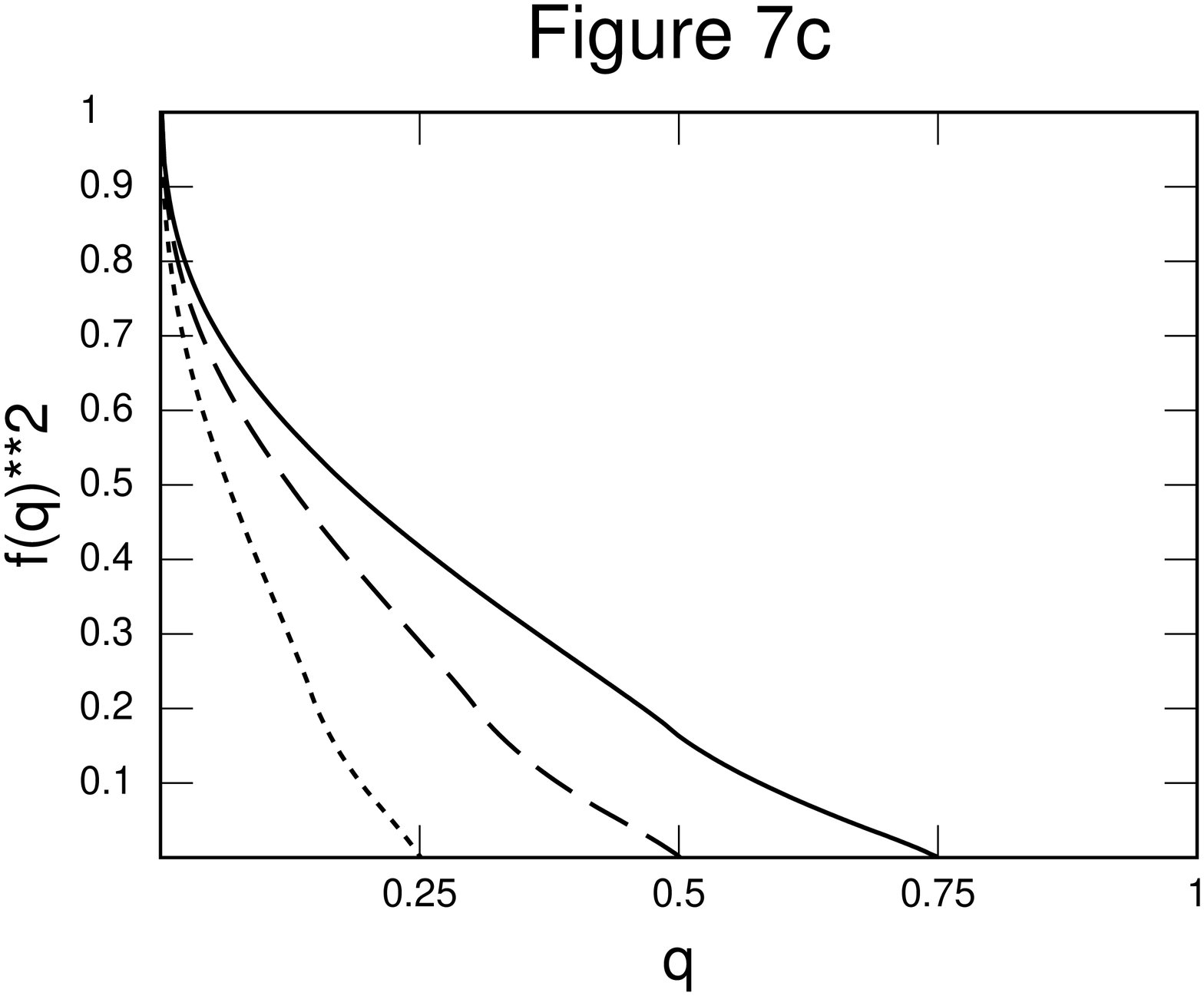}}
\end{figure}

\vspace*{4in}

\newpage

~

\begin{figure}
\epsfxsize 4.0 in
\mbox{\epsfbox{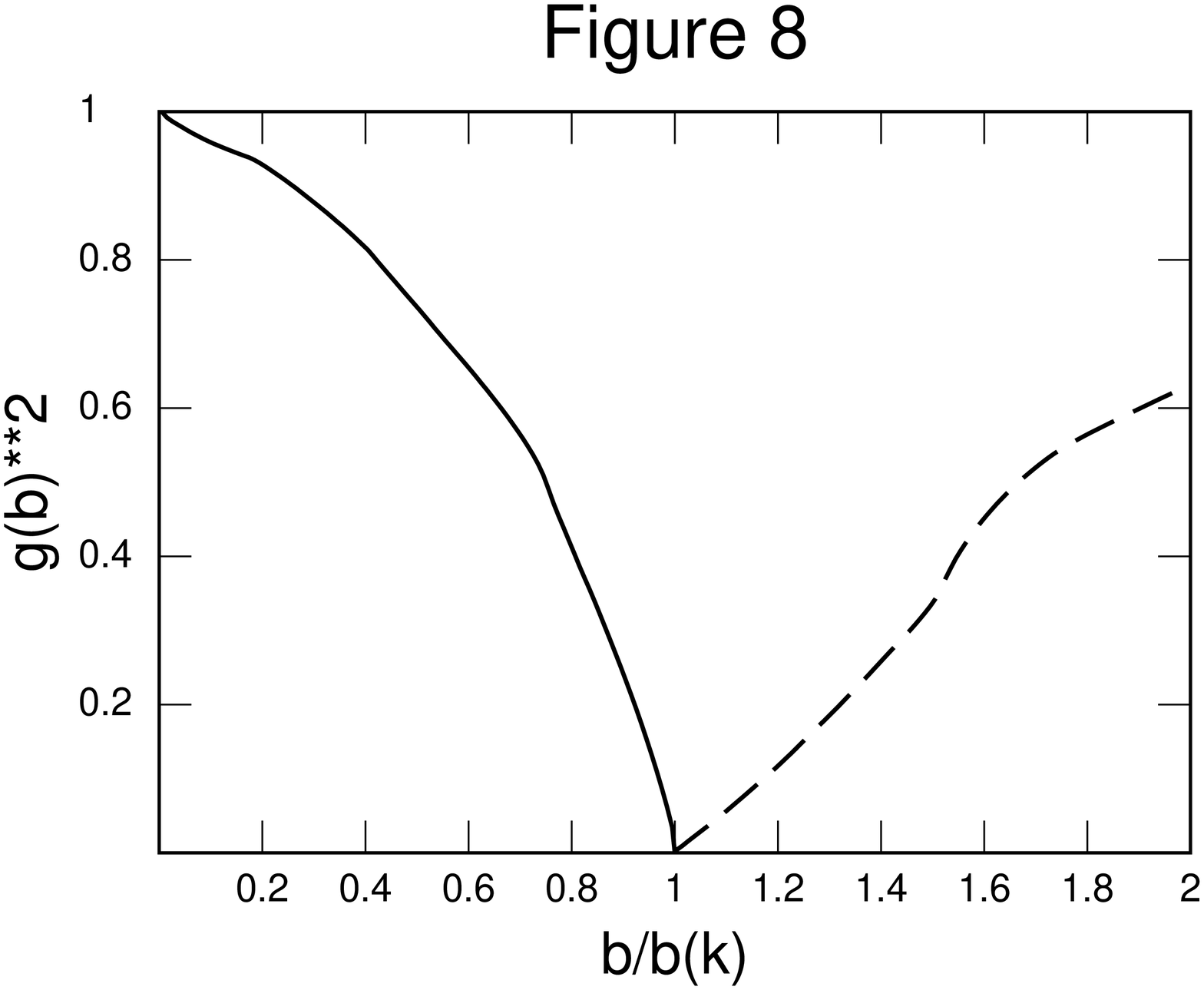}}
\end{figure}

\end{document}